# Coherence properties of the high-energy fourth-generation X-ray synchrotron sources

R. Khubbutdinov[1,2], A. P. Menushenkov[2], and I. A. Vartanyants [1,2,*]

[1]*Deutsches Elektronen-Synchrotron DESY, Notkestrasse 85, D-22607 Hamburg, Germany*

[2]*National Research Nuclear University MEPhI (Moscow Engineering Physics Institute)*

*Kashirskoe shosse 31, 115409 Moscow, Russia*

July 8, 2019

We performed an analysis of coherence properties of the 4[th] generation high-energy storage rings with emittance values of 10 pmrad. It is presently expected that a storage ring with these low emittance values will reach diffraction limit at hard X-rays. Simulations of coherence properties were performed with the XRT software and analytical approach for different photon energies from 500 eV to 50 keV. It was demonstrated that a minimum photon emittance (diffraction limit) reached at such storage rings is $\lambda/2\pi$. Using mode decomposition we showed that at the parameters of the storage ring considered in this work, diffraction limit will be reached for soft X-ray energies of 500 eV. About ten modes will contribute to the radiation field at 12 keV photon energy and even more modes give a contribution at higher photon energies. Energy spread effects of the electron beam in a low emittance storage ring were analysed in detail. Simulations were performed at different relative energy spread values from zero to $2 \cdot 10^{-3}$. We observed a decrease of the degree of coherence with an increase of the relative energy spread value. Our analysis shows that to reach diffraction limit for high photon energies electron beam emittance should go down to 1 pmrad and below.

---
[*] corresponding author: Ivan.Vartaniants@desy.de

## 1. Introduction

Recently it was realized, that due to the availability of new technology, *i.e.* multi-bend achromat synchrotron storage ring design, the brilliance of next generation x-ray storage rings may be increased by two-three orders of magnitude (Eriksson *et al.*, 2014; Hettel, 2014). This immediately implies that coherent flux of these storage rings will be as well by two orders of magnitude higher and, by that, will be approaching the so-called diffraction limit. Next generation synchrotron sources have the potential to make a great contribution to today's major challenges in investigation of multi-functional hybrid materials, electronic transport phenomena and electrochemical processes in charge storage materials under working conditions, as well as materials under extreme conditions of pressure and temperature with highest resolution and sensitivity (Weckert, 2015).

The first storage ring constructed using multi-bend achromat technology was 3 GeV synchrotron source MAX IV (Lund, Sweden), which recently reached its planned specifications of horizontal emittance of about 200 - 330 pmrad (depending on insertion device gap settings) (Tavares *et al.*, 2014). The high-energy ESRF 6 GeV storage ring is presently under reconstruction to EBS ESRF facility (EBS ESRF, n.d.) that is planned to reach horizontal emittance of 133 pmrad. Brazilian SIRIUS 3 GeV project is presently in the commissioning phase with horizontal emittance planned in the range of 150 – 250 pmrad (Rodrigues *et al.*, 2018) and other facilities worldwide (APS-U, SPring-8, ALS, Soleil, Diamond, and *etc*.) are in construction and planning stage. At DESY in Hamburg an upgrade of high-energy 6 GeV storage ring PETRA III to PETRA IV facility is also planned (Schroer *et al.*, 2019). The world lowest emittance of about 10 pmrad for hard x-rays is targeted at this storage ring (Schroer *et al.*, 2018).

Source brilliance and coherence of the future storage rings are the keys for successful synchrotron radiation experiments. High values of these properties will allow the focusing of the synchrotron beams efficiently to the nanometer range without spatial filtering of flux (Singer & Vartanyants, 2014). It will allow an effective application of coherence based techniques such as Coherent Diffraction Imaging (CDI) potentially reaching sub-nanometer resolution (Schroer & Falkenberg, 2014). It will also extend photon correlation techniques into the regime of nanoseconds and allow for low dose correlation experiments (Shpyrko, 2014).

In order to reach all these goals, a better understanding of the coherence properties of radiation close to the diffraction limit is necessary. Ultimate storage rings are expected to have a high degree of coherence, which means that traditional methods of X-ray tracing will not be sufficient to predict parameters of X-ray beams at the experimental stations. Also, some intrinsic electron beam parameters such as an unavoidable energy spread of electrons in a storage ring may influence the coherence properties of X-ray beams.

It is interesting to note that even basic parameters of diffraction-limited sources are not well understood nowadays. For example, it is a long-standing debate on what is a correct asymptotic limit for the photon emittance of a diffraction-limited storage ring (Walker, 2019). It is commonly accepted that the diffraction-limited storage ring is the one with its electron beam emittance much lower than the natural emittance of single electron radiation. For radiation described by the Gaussian functions, the diffraction-limited emittance is very well known from the uncertainty principle and is given by $\lambda/4\pi$ (Kim, 1989). At the same time, several authors have demonstrated that the emittance of a single electron undulator radiation is given by $\lambda/2\pi$ (Onuki & Elleaume, 2003; Tanaka & Kitamura, 2009). The problem is based on the fact that radiation from a single electron cannot be described by a Gaussian function. It is well known that in the far-field region and resonant conditions it is described by a sinc-function (Alferov *et al.*, 1973; Kim, 1986). Further on, we will carefully analyze this question and will give a definitive answer to this long-standing debate[1].

Another important question is the influence of the energy spread of electrons in a storage ring on coherence properties of the photon beams. It was first outlined by Tanaka&Kitamura  (Tanaka & Kitamura, 2009) that energy spread effects may affect brilliance of the X-ray storage rings. In a recent paper by Geloni et al., (Geloni *et al.*, 2018) energy spread effects on brightness and coherence were carefully analysed for the low-emittance storage rings using an analytical approach. As a result of this analysis, it was shown that brightness of diffraction limited sources with the small but finite emittance only slightly depends on the energy spread effects, the same is valid for coherence function centered at the optical axis of the undulator with stronger effects at off-axis values. It was also demonstrated that the values of brightness obtained by this approach substantially differ from the results of Tanaka & Kitamura (Tanaka & Kitamura, 2009). We will discuss in detail an effect of energy spread on the coherence properties of the low-emittance storage rings.

---

[1] When this work was finished the paper written by Walker (Walker, 2019) was published. The thorough analysis presented by the author was based on the synchrotron radiation simulations presented in the book (Onuki & Elleaume, 2003). The author also discussed the question of diffraction limit and came to similar conclusion as in our work for the so-called projected values of emittance.

Presently it is a big demand for developing adequate and effective methods that may correctly describe properties of radiation from the ultimate storage rings close to the diffraction limit. Clearly, such a description should be based on the application of the first- and higher-order correlation functions (Mandel & Wolf, 1995; Vartanyants & Singer, 2016). Presently several codes were developed to describe partial coherence radiation from conventional storage rings (see, for example, (Lee & Zhang, 2007; Shi *et al.*, 2014; Chubar & Elleaume, 1998; Meng *et al.*, 2015)). Unfortunately, these codes may be inefficient for simulating properties of diffraction-limited storage rings or require a substantial computer power and time to perform simulations. In this work for the analysis of coherence properties of the low-emittance storage rings, we used recently developed computer code XRT (Klementiev & Chernikov, 2014) that allows efficiently simulating all necessary correlation functions.

In this work, we analysed coherence properties of a high-energy low-emittance (10 pmrad and below) storage ring for different photon energies starting from soft X-rays of 500 eV up to hard X-rays of 50 keV. We introduced energy spread effects in our simulations and specifically examined the following values: zero energy spread value, $1 \cdot 10^3$, and $2 \cdot 10^3$. These studies may be of particular interest for the future PETRA IV facility with its record parameters (Schroer *et al.*, 2018).

The paper is organized as follows. In the next section, a short theoretical summary of the first-order correlation functions will be presented. In the same section, the basic principles of undulator radiation and its analytical description will be discussed. In the third and main section of the paper, simulations and results of the analysis of coherence properties of the low-emittance storage ring will be performed by different analysis tools. The paper will be finished with conclusions and outlook.

## 2. Theory

### *a) Basics of coherence theory*

The measure of the first-order coherence is given by the mutual coherence function (MCF) defined as (Mandel & Wolf, 1995)

$$\Gamma(\boldsymbol{r_1}, \boldsymbol{r_2}, t_1, t_2) = <E^*(\boldsymbol{r_1}, t_1)E(\boldsymbol{r_2}, t_2)> . \qquad (1)$$

It describes correlations between two values of the electric field $E(\boldsymbol{r_1}, t_1)$ and $E(\boldsymbol{r_2}, t_2)$ at different points $\boldsymbol{r_1}$ and $\boldsymbol{r_2}$ and times $t_1$ and $t_2$. The brackets <...> denote ensemble average. We next introduce the cross-spectral density function (CSD) of a stationary source, which is obtained as the Fourier transform of the MCF

$$W(\boldsymbol{r_1}, \boldsymbol{r_2}, \omega) = \int \Gamma(\boldsymbol{r_1}, \boldsymbol{r_2}, \tau)\, e^{-i\omega\tau} d\tau , \qquad (2)$$

where $\tau = t_2 - t_1$ . The spectral density of the radiation field is obtained when two points $\boldsymbol{r_1}$ and $\boldsymbol{r_2}$ coincide $\boldsymbol{r} = \boldsymbol{r_1} = \boldsymbol{r_2}$

$$\mathrm{S}(\boldsymbol{r}, \omega) = \mathrm{W}(\boldsymbol{r}, \boldsymbol{r}, \omega). \qquad (3)$$

A convenient measure of spatial coherence is the normalized CSD

$$\mu(\boldsymbol{r_1}, \boldsymbol{r_2}, \omega) = \frac{W(\boldsymbol{r_1}, \boldsymbol{r_2}, \omega)}{\sqrt{S(\boldsymbol{r_1}, \omega)}\sqrt{S(\boldsymbol{r_2}, \omega)}} , \qquad (4)$$

which is called the spectral degree of coherence (SDC). The values of this function, which are ranging from zero to one and depend on the pinhole separation $\boldsymbol{r_1}$ and $\boldsymbol{r_2}$, are determined in the classical Young's experiment.

Another convenient measure of coherence is the global degree of coherence $\zeta^{DC}$, which characterizes coherence properties of the wavefield by a single number and can be introduced as (Geloni *et al.*, 2008; Vartanyants & Singer, 2010)

$$\zeta^{DC}(\omega) = \frac{\int |W(\boldsymbol{r_1}, \boldsymbol{r_2}, \omega)|^2 d\boldsymbol{r_1} d\boldsymbol{r_2}}{(\int S(\boldsymbol{r}, \omega) d\boldsymbol{r})^2} . \qquad (5)$$

The values of the parameter $\zeta^{DC}(\omega)$ lie in the range of $0 \leq \zeta^{DC}(\omega) \leq 1$, where $\zeta^{DC}(\omega) = 1$ and $\zeta^{DC}(\omega) = 0$ correspond to fully coherent and incoherent radiation, respectively.

In the quasi-monochromatic regime, it is possible to approximate the MCF in Eq. (1) as

$$\Gamma(\boldsymbol{r_1}, \boldsymbol{r_2}, \tau) \approx J(\boldsymbol{r_1}, \boldsymbol{r_2})e^{-i\omega\tau}, \tag{6}$$

provided that $|\tau| \ll 1/\Delta\omega$, where $\Delta\omega$ is the bandwidth of radiation. Here $J(\boldsymbol{r_1}, \boldsymbol{r_2})$ is the Mutual Optical Intensity (MOI) defined as

$$J(\boldsymbol{r_1}, \boldsymbol{r_2}) = <E^*(\boldsymbol{r_1}, t)E(\boldsymbol{r_2}, t)>. \tag{7}$$

Taking into account all of the above, we may represent CSD for quasi-monochromatic radiation as

$$W(\boldsymbol{r_1}, \boldsymbol{r_2}, \omega) = J(\boldsymbol{r_1}, \boldsymbol{r_2})\delta(\omega - \omega_0), \tag{8}$$

which describes correlations between two complex values of the electric field at different points $\boldsymbol{r_1}$ and $\boldsymbol{r_2}$ at a given frequency.

Finally, for quasi-monochromatic radiation CSD $W(\boldsymbol{r_1}, \boldsymbol{r_2}, \omega)$ and MOI $J(\boldsymbol{r_1}, \boldsymbol{r_2})$ functions as well as spectral density $S(\boldsymbol{r}, \omega)$ and intensity $I(\boldsymbol{r})$ functions are equivalent.

### b) Coherent-mode representation of the cross-spectral density function

It is well known (Mandel & Wolf, 1995) that, under very general conditions, one can represent the CSD of a partially coherent, statistically stationary field of any state of coherence as a series

$$W(\boldsymbol{r_1}, \boldsymbol{r_2}, \omega) = \sum_j \beta_j(\omega)E_j^*(\boldsymbol{r_1}, \omega)E_j(\boldsymbol{r_2}, \omega). \tag{9}$$

Here $\beta_j(\omega)$ are eigenvalues and independent coherent modes $E_j(\boldsymbol{r}, \omega)$ are eigen-functions of the Fredholm integral equation of the second kind

$$\int W(\boldsymbol{r_1}, \boldsymbol{r_2}, \omega)E_j(\boldsymbol{r_1}, \omega)d\boldsymbol{r_1} = \beta_j(\omega)E_j(\boldsymbol{r_2}, \omega). \tag{10}$$

According to Eq. (3) and Eq. (9) the spectral density can be represented as

$$S(\mathbf{r}, \omega) = \sum_j \beta_j(\omega)\left|E_j(\boldsymbol{r}, \omega)\right|^2. \tag{11}$$

Substitution of Eq. (9) and Eq. (11) into Eq. (5) gives for the global degree of coherence

$$\zeta^{DC}(\omega) = \frac{\sum \beta_j^2(\omega)}{\left(\sum \beta_j(\omega)\right)^2} \quad . \tag{12}$$

One has to remember important characteristics of this coherent mode decomposition: the mode functions $E_j(\boldsymbol{r}, \omega)$ form an orthonormal set, the eigenvalues $\beta_j(\omega)$ are real and non-negative $\beta_j(\omega) \geq 0$, and $\beta_0(\omega) \geq \beta_1(\omega) \geq \dots$ . If there is only one single mode present then radiation is fully coherent. Thus we can define the coherent fraction (CF) of radiation $\zeta^{CF}(\omega)$ as an occupation or normalized weight of the first mode

$$\zeta^{CF}(\omega) = \frac{\beta_0(\omega)}{\sum_{j=0}^{\infty} \beta_j(\omega)}. \tag{13}$$

We will consider a quasi-monochromatic case and will omit frequency dependence in the following.

### c) *Gaussian Schell – model sources*

Gaussian Schell-model (GSM) is a simplified but often used model (Vartanyants & Singer, 2010) that represents radiation from a real X-ray source based on the following approximations. The source is modelled as a plane two-dimensional source, the source is spatially uniform, i.e. the SDC depends only on the difference $\boldsymbol{r_2} - \boldsymbol{r_1}$, the SDC $\mu(\boldsymbol{r_2} - \boldsymbol{r_1})$ and spectral density $S(\boldsymbol{r})$ are Gaussian functions. In the frame of GSM cross-spectral density function, spectral density and SDC are defined as (Mandel & Wolf, 1995)

$$W(\boldsymbol{r_1}, \boldsymbol{r_2}) = \mu(\boldsymbol{r_2} - \boldsymbol{r_1})\sqrt{S(\boldsymbol{r_1})}\sqrt{S(\boldsymbol{r_2})} \ , \tag{14}$$

$$S(\boldsymbol{r}) = S_0 \exp\left(-\frac{r_x^2}{2\sigma_x^2} - \frac{r_y^2}{2\sigma_y^2}\right), \tag{15}$$

$$\mu(\boldsymbol{r_2} - \boldsymbol{r_1}) = \exp\left(-\frac{(r_{x2} - r_{x1})^2}{2\xi_x^2} - \frac{(r_{y2} - r_{y1})^2}{2\xi_y^2}\right), \tag{16}$$

where $S_0$ is a normalization constant, $\sigma_{x,y}$ is the rms source size and $\xi_{x,y}$ is the transverse coherence length in the source plane in *x*- and *y*- direction, respectively. One of the important features of this model is that CSD function is separable into two transverse directions

$$W(\boldsymbol{r_1}, \boldsymbol{r_2}) = W(r_{x1}, r_{x2})W(r_{y1}, r_{y2}) \ . \tag{17}$$

The same is valid for the global degree of coherence defined in Eq. (5)

$$\zeta^{DC} = \zeta_x^{DC}\zeta_y^{DC} , \qquad (18)$$

where in each transverse direction $i = x, y$ we have (Vartanyants & Singer, 2010)

$$\zeta_i^{DC} = \frac{\int |W(r_{i1}, r_{i2})|^2 dr_{i1} dr_{i2}}{(\int S(r_i) dr_i)^2} = \frac{\xi_i/\sigma_i}{\sqrt{\left(\frac{\xi_i}{\sigma_i}\right)^2 + 4}} . \qquad (19)$$

Coherent modes in the GSM are described by the Hermite-Gaussian functions (Gori, 1983; Starikov & Wolf, 1982; Vartanyants & Singer, 2010)

$$\beta_j/\beta_0 = \varkappa^j , \qquad (20)$$

$$E_j(r_i) = \left(\frac{\varkappa_i}{2\pi k \sigma^2 \zeta_i}\right)^{1/4} \frac{1}{\sqrt{2^j j!}} H_j\left(\frac{r_i}{\sigma\sqrt{2\zeta_i}}\right) \exp\left(-\frac{r_i^2}{4\sigma^2 \zeta_i}\right), \qquad (21)$$

where the coefficient $\varkappa_i = (1 - \zeta_i)/(1 + \zeta_i)$ is introduced, $H_j\left(r_i/\sigma\sqrt{2\zeta_i}\right)$ are the Hermite polynomials of order $j$, $k = 2\pi/\lambda$ and $\lambda$ is the wavelength of radiation, and $i = x, y$. The zero mode is a Gaussian function and propagation of Hermite-Gaussian modes in the far-field region gives Hermite-Gaussian modes of the same shape. In the frame of GSM, according to Eqs. ((13, (20) coherent fraction of the radiation for one transverse direction may be determined as

$$\zeta_i^{CF} = \left(\sum_{j=0}^{\infty} \frac{\beta_j(\omega)}{\beta_0(\omega)}\right)^{-1} = \left(\sum_{j=0}^{\infty} \varkappa^j\right)^{-1} \approx \frac{2\zeta_i}{1 + \zeta_i} . \qquad (22)$$

### 3.    Synchrotron radiation

a) *Spectral brightness and phase space distribution*

The source may be completely characterized by its spectral brightness which is defined through the phase space distribution function that is a classical analogue of the Wigner distribution function (Wigner, 1932). According to this definition, the spectral brightness $B(\boldsymbol{r}, \boldsymbol{k_\perp})$ in paraxial approximation is given by (Mandel & Wolf, 1995)

$$B(\boldsymbol{r}, \boldsymbol{k_\perp}) = \left(\frac{k}{2\pi}\right)^2 \int W_0\left(\boldsymbol{r} - \frac{1}{2}\Delta\boldsymbol{r}, \boldsymbol{r} + \frac{1}{2}\Delta\boldsymbol{r}\right) e^{i\boldsymbol{k_\perp}\cdot\Delta\boldsymbol{r}} d\Delta\boldsymbol{r} , \qquad (23)$$

where $W_0\left(r - \frac{1}{2}\Delta r, r + \frac{1}{2}\Delta r\right)$ is the CSD function (2,8) defined at the source position, $k_\perp$ is projection of the momentum vector $k$ on the transverse plane, and the coordinates $r$ and $\Delta r$ are introduced as $r = (r_1 + r_2)/2$ and $\Delta r = (r_1 - r_2)$.

In the synchrotron radiation community it is conventional to define the distribution function in Eq. (23) in the phase space through the CSD function $W_0(r, \Delta r)$ of the electric fields at the source position (Kim, 1989; Geloni *et al.*, 2015)

$$B(r, \theta) = \frac{c}{(2\pi)^4} \frac{I}{e\hbar} k^2 \int W_0(r, \Delta r) \exp(-ik\theta \cdot \Delta r)\, d\Delta r , \qquad (24)$$

where $\theta$ is the angle between the optical axes and observation point, $I$ is the electron beam current, $e$ is an electron charge, and $\hbar = h/2\pi$ is the Planck's constant. This distribution taken at its maximum value, which is typically for undulator radiation coincides with the optical axis for odd harmonics, is defined as spectral brightness of the synchrotron source.

Approximating far-field distribution of the electric field of undulator radiation in resonant conditions from a single electron by a Gaussian laser mode the following well known approximation for the brightness of undulator radiation was obtained (Kim, 1986; Kim, 1989)

$$B_0 = B(0,0) = \frac{1}{4\pi^2} \frac{F}{\Sigma_{ph\,x} \Sigma'_{ph\,x} \Sigma_{ph\,y} \Sigma'_{ph\,y}} , \qquad (25)$$

where $F$ is the spectral photon flux into the central core. In Eq. (25) total photon source size $\Sigma_{ph}$ and divergence $\Sigma'_{ph}$ are defined as

$$\Sigma_{ph} \approx \left[\sigma_e^2 + \sigma_r^2\right]^{\frac{1}{2}}, \qquad \Sigma'_{ph} \approx \left[\sigma_e'^2 + \sigma_r'^2\right]^{\frac{1}{2}}, \qquad (26)$$

where $\sigma_e = \sqrt{\varepsilon_e \beta_e}$ and $\sigma_e' = \sqrt{\varepsilon_e / \beta_e}$ are the electron beam spatial size and angular divergence, $\varepsilon_e = \sigma_e \sigma_e'$ is the emittance of an electron beam, and $\beta_e$ is the value of the betatron function in the center of the undulator. Intrinsic characteristics of a single-electron radiation in Eq. (26) are introduced in (Kim, 1989)

$$\sigma_r = \frac{\sqrt{2\lambda_n L_u}}{4\pi} , \qquad \sigma_r' = \sqrt{\frac{\lambda_n}{2L_u}} , \qquad (27)$$

where $\lambda_n$ and $L_u$ are the $n$-th harmonic radiation wavelength and the undulator length. The total photon emittance of the undulator source is then introduced as

$$\varepsilon_{ph\,x,y} = \Sigma_{ph\,x,y} \cdot \Sigma'_{ph\,x,y}. \tag{28}$$

By that approach, the photon emittance $\varepsilon_r$ or photon phase space of a single electron in Gaussian approximation is given by the value (Kim, 1989)

$$\varepsilon_r = \sigma_r \sigma'_r = \lambda/4\pi. \tag{29}$$

The diffraction-limited source may be defined as the one with the electron beam parameters satisfying the following inequalities: $\sigma_e \ll \sigma_r$ and $\sigma'_e \ll \sigma'_r$, According to these definitions for the diffraction-limited source the natural electron beam emittance has to be much smaller than natural emittance of a single electron radiation $\varepsilon_e \ll \varepsilon_r \equiv \lambda/4\pi$.

b) *Energy spread effects*

The energy spread of the electron beam is an energy distribution of electrons in the bunch, which for synchrotron radiation obeys Gaussian statistics. It was suggested by Tanaka & Kitamura (Tanaka & Kitamura, 2009) that energy spread effects may influence properties of the synchrotron radiation source. Approximating the angular and spatial profile of the flux density by Gaussian functions for each transverse direction the following expressions for the beam size and divergence were obtained (Tanaka & Kitamura, 2009)

$$\Sigma_{ph} \approx \left[\varepsilon\beta + \sigma_r^2 * 4 * N\left(\frac{\upsilon}{4}\right)^{\frac{2}{3}}\right]^{\frac{1}{2}}, \Sigma'_{ph} \approx \left[\frac{\varepsilon}{\beta} + \sigma_r'^2 * N(\upsilon)\right]^{\frac{1}{2}}, \tag{30}$$

where $\sigma_r$ and $\sigma'_r$ are intrinsic characteristics of a single-electron radiation, defined in Eqs. (27). A normalization factor $N(\upsilon)$ for the energy spread of the storage ring in Eqs. (30) is defined as

$$N(\upsilon) = \frac{8\pi^2\upsilon^2}{(2\pi)^{3/2}\upsilon\,\mathrm{erf}\left[(8\pi^2)^{\frac{1}{2}}\upsilon\right] + \exp(-8\pi^2\upsilon) - 1}, \tag{31}$$

with $\upsilon = \delta_\gamma/\delta_n$ being the ratio between the relative energy spread value $\delta_\gamma = \Delta E/E$ and the relative bandwidth $\delta_n = 1/nN_u$ of the $n^{\mathrm{th}}$ harmonic of an undulator with $N_u$ periods. Normalization function $N(\upsilon)$ for zero relative energy spread is equal to one $N(0) = 1$. Note, that the source size for natural single-electron radiation in Eq. (30) is effectively two times larger than defined in Eqs. (27). This difference originates from a non-Gaussian angular profile of far-field radiation from a single electron that is taken into account in Eq. (30). This also leads to a different condition for the diffraction-limited source: $\varepsilon_e \ll \varepsilon_r = 2\sigma_r\sigma'_r \equiv$ $\lambda/2\pi$. We will see in the following by performing simulations which of two condition is satisfied in the limit of small electron beam emittance for the diffraction-limited source.

In a recent paper by Geloni et al., (Geloni *et al.*, 2018) energy spread effects on brightness and coherence were carefully analysed using directly Eq. (24) with an account of energy spread effects. For the far-field distribution of the electric field with a contribution of the energy spread of relativistic electron subjected to periodic acceleration in an undulator, the following expression was used in Eq. (24) (Geloni et al., 2018)

$$E(\boldsymbol{\theta}) = -\frac{K\omega e L_u}{2c^2 z \gamma} A_{jj} exp\left[i\frac{\omega}{c}\left(\frac{z\boldsymbol{\theta}^2}{2} - \boldsymbol{\theta}\boldsymbol{l}\right)\right] sinc\left[\frac{2\pi N_u(\gamma - \gamma_1)}{\gamma_1} + \frac{\omega L_u|\boldsymbol{\theta} - \boldsymbol{\eta}|^2}{4c}\right], \tag{32}$$

where $\gamma$ and $\gamma_1$ are the Lorentz factors at a given and resonant frequency electron energy, z is the distance from the center of the undulator source, $K$ is the maximum undulator deflection parameter, and the coupling parameter $A_{jj}$ is defined as

$$A_{jj} = J_0\left(\frac{K^2}{4 + 2K^2}\right) - J_1\left(\frac{K^2}{4 + 2K^2}\right),$$

where $J_n$ being the n[th] order Bessel function of the first kind. In expression (32) energy offset $\Delta\gamma$, different entering angles $\boldsymbol{\eta}$ and various axis offset $\boldsymbol{l}$ of the electron in the undulator are explicitly taken into account.

In our further analysis we will need the amplitude of the field at the source position. It is determined directly from expression (32) by applying the propagator in free space in paraxial approximation and is given by the following expression (Geloni *et al.*, 2015)

$$E_0(\boldsymbol{r}) = \frac{iz\omega}{2\pi c} \int E(\boldsymbol{\theta}) exp\left(-\frac{i\theta^2 z\omega}{2c}\right) exp\left(\frac{i\omega \boldsymbol{r}\boldsymbol{\theta}}{c}\right) d\boldsymbol{\theta} . \tag{33}$$

To obtain the source size and divergence for non-Gaussian field distributions we determined them by calculating the second moments or variances of the intensity distribution of the corresponding variables in each direction, respectively (Onuki & Elleaume, 2003)

$$\Sigma^2{}_{ph\,x} = \frac{\iint x^2 |E_0(x,y)|^2 dxdy}{\iint |E_0(x,y)|^2 dxdy} , \qquad \Sigma'^2{}_{ph\,x} = \frac{\iint \theta_x{}^2 \left|E(\theta_x,\theta_y)\right|^2 d\theta_x d\theta_y}{\iint \left|E(\theta_x,\theta_y)\right|^2 d\theta_x d\theta_y}. \tag{34}$$

In addition to energy spread effects, the source size and source divergence are influenced by undulator detuning – a shift of the undulator harmonic wavelength $\lambda_n$ relative to the selected radiation wavelength λ (Coisson, 1988). Detuning may happen as a result of improper synchronization between the monochromator and the insertion device on a beamline. Depending on its value, detuning may have a stronger effect (a bigger linear and angular size variation) than energy spread but is out of scope of the present paper.

### c) *Coherent fraction*

It is well established in the synchrotron radiation community that coherent flux of a zero-emittance beam is given by $F_{coh} = (\lambda/2)^2 B_0$ (see, for example, Kim, 1986; Geloni, 2015). Importantly, this is an exact result that does not depend on the Gaussian approximation of single-electron radiation. Next, if we define the flux of synchrotron source with certain values of emittance through brightness as in Eq. (25), we obtain

$$\text{F} = 4\pi^2 \varepsilon_{ph\,x} \varepsilon_{ph\,y} B_0. \tag{35}$$

Strictly speaking, this expression is valid only for a Gaussian approximation of the electron-beam source parameters and natural single electron radiation with emittance values $\varepsilon_{ph\,x,y}$ defined in Eq. (28) (see, for discussion, Geloni , 2018). We will introduce coherent fraction as

$$\zeta^{CF} = \frac{F_{coh}}{F} = \left(\frac{\lambda}{4\pi}\right)^2 \frac{1}{\varepsilon_{ph\,x} \varepsilon_{ph\,y}} = \frac{\varepsilon_r^2}{\varepsilon_{ph}^x \varepsilon_{ph}^y}, \tag{36}$$

where $\varepsilon_r$ is the conventional emittance of single-electron radiation defined in Eq. (29). We will see in the following how expression (36) will be modified due to the non-Gaussian behavior of single-electron radiation in the far-field region.

## 4.    Results and discussion

As an example of the diffraction-limited source, we considered high-energy storage ring operating at 6 GeV (for parameters of the source see Table 1 and Table 2). For the electron emittance as a basic value we considered 10 pmrad both in the vertical and horizontal direction by that considering rather a round beam shape. We also analysed a broader range of electron emittance values from 1 pmrad to 300 pmrad to have a better understanding of the photon properties of the source. Simulations were performed either in the far-field region at the distance of 30 m from the source or directly at the source position that was considered at the middle of the undulator. Simulations were performed for the four energy values of 500 eV, 12 keV, 24 keV, and 50 keV. Comparing natural emittance of single-electron radiation at these energies (see Table 2) with the designed parameters of the source we see that at energy of 12 keV emittance values are comparable, at 500 eV $\varepsilon_e \ll \varepsilon_r$ and at 24 keV as well as at 50 keV $\varepsilon_e > \varepsilon_r$. From that, we may expect to reach the diffraction-limited source at 500 eV and have parameters of radiation close to the diffraction-limited source at 12 keV. We also would expect that at higher energies of 24 keV and 50 keV radiation will be highly coherent but not diffraction limited. All simulations were performed with the XRT software (Klementiev & Chernikov, 2014) and compared with the analytical approach.

### a)   Photon emittance

Results of simulations performed by the XRT software of the photon emittance as a function of the electron beam emittance values from 1 pmrad to 300 pmrad for different relative energy spread values are presented in Fig. 1. X-ray radiation of the electron beam was simulated for the synchrotron storage ring with the parameters presented in the Table 1 and Table 2 (see details in the supporting information). The photon emittance (triangles in Fig. 1) was simulated as a product of the source size and divergence (see Eq. (28)) which were obtained as the variance values of the intensity distributions of the corresponding variables at the source position or in the far-field region according to Eq. (34).

As a result of these simulations, we can see the lower is electron emittance $\varepsilon_e$ the lower is the photon emittance $\varepsilon_{ph}$. Importantly, we can observe that at 500 eV and 12 keV photon energies and at 10 pmrad electron beam emittance, the photon emittance reaches its asymptotic value (note also different scale for 500 eV photon energy). This is a clear indication that at these energies the synchrotron source may be considered as diffraction limited. We also observe that at larger energy spread values and the same electron emittance, photon emittance is also increasing. However, energy spread induced difference does not exceed 12% at 500 eV/10 pmrad, while at higher energy this difference goes up to 50 %. This is due to the fact that at low photon energies the properties of the beam (source size and divergence) are comparably large, and their small changes caused by the energy spread effect are not noticeable. In contrast, the energy spread effects are revealed strongly at high energies due to smaller parameters of the radiation at these energies.

We compared these results with the analytical ones obtained using Eqs. (32-34) (circles in Fig. 1) (see for details the supporting information). We see that the results of the analytical approach and simulations performed with the XRT software correspond to each other very well (both for different electron emittance values and for different energy spread values) within the margins of the error bars. Both approaches were also compared with the calculations made according to the approach of Tanaka&Kitamura (Tanaka & Kitamura, 2009) Eqs. (30-31) (shown by lines in Fig. 1). We observe that all three approaches give similar results for three values of energy spread that were considered here.

As can be clearly seen in Fig. 1 (see also Table 3), the lowest value of the photon emittance for zero energy spread in our simulations is asymptotically reaching the value of $\lambda/2\pi$, when electron emittance values are becoming sufficiently small. This is a strong indication that for low-emittance storage rings X-ray radiation cannot be approximated as Gaussian, because in this case the lowest photon emittance should reach the value of $\lambda/4\pi$ as suggested in Ref. (Kim, 1989). At the same time, these results are in concordance with the results of Ref. (Tanaka & Kitamura, 2009) where non-Gaussian behaviour of synchrotron radiation of a single electron was analysed.

Now we would like to determine a coherent fraction of radiation at different electron emittance values. If we will use conventional expression (36) with the emittance values $\varepsilon_{ph}$ shown in Fig. 1 our coherent fraction values would reach an asymptotic value of 0.5 and would never reach one. We slightly redefined our previous expression to the new one (see also Onuki & Elleaume, 2003)

$$\zeta^{CF} = \frac{F_{coh}}{F} = \frac{\varepsilon_{coh}^2}{\varepsilon_{ph}^x \varepsilon_{ph}^y}, \tag{37}$$

where $\varepsilon_{coh} = \lambda/2\pi$ and $\varepsilon_{ph}$ are photon emittance values obtained through different simulations.

The results of our simulations in one transverse direction are presented in Fig. 2. First, we determined coherent fraction from XRT simulations (triangles) by using expression (37) and the results of emittance simulations shown in Fig. 1. Then we compared results of XRT simulations with the analytical calculation. Expressions (32) and (33) were used to calculate the wave field amplitudes in the far-field region. Eqs. (34) were applied to determine the source parameters. Using the same expression (37) we obtained analytical values (circles) of coherent fraction shown in Fig. 2. Finally, we used expressions (30) and (31) of Tanaka&Kitamura (Tanaka & Kitamura, 2009) in equation (37) (lines).

We see that all three results excellently agree with each other and show the same trend that was just discussed for emittance. It is important to note, that to obtain this result we have to use expression (37) instead of commonly used Eq. (36). Next, we will turn to a more general definition of coherent fraction through coherent mode decomposition.

b) *Representation of the cross-spectral density by the coherent mode decomposition*

As it was discussed in the Theory section the four-dimensional CSD function may be decomposed to a sum of two-dimensional coherent modes as given in Eq. (9). The XRT software is providing an opportunity to directly analyze these modes.

We performed mode decomposition of the CSD according to Eqs. (9-11) using the XRT software. We performed our analysis for the studied case of 10 pmrad electron beam emittance. The mode decomposition was used to determine the shape and contribution of each mode at different photon energies and at different values of relative energy spread. The first four modes and their normalized weights for 500 eV and 12 keV photon energy are shown in Fig. 3 (results of simulations for 24 keV and 50 keV are given in the supporting information Fig. S2). An orthogonal set of modes determined by our simulations, for the whole range of energy spread values from 0 to $2 \cdot 10^{-3}$, represents a mixture of Hermite-Gaussian and Laguerre-Gaussian modes (Siegman, 1986). These two different sets of modes have different symmetry: Hermite-Gaussian modes may be represented as a product of two separable amplitude functions in the transverse plane and, contrary to that, Laguerre-Gaussian modes possess cylindrical symmetry and do not allow decomposition in two orthogonal directions. It can be also noticed, that the contribution of Laguerre-Gaussian modes is increased for both energies with the increase of the energy spread values. It is interesting to note here that the contribution of Laguerre-Gaussian modes is coming solely from the energy spread effect. It can be shown, that if all electron offset parameters (entering angle $\boldsymbol{\eta}$ and axis offset $\boldsymbol{l}$) are put to zero then all modes of radiation field may be represented by Laguerre-Gaussian modes. This all means that, in general, for the diffraction-limited source, it is not possible to factorize CSD function in two orthogonal transverse directions and define the degree of coherence as a product of its values in each direction (see Eq. (18)).

We also estimated the number of modes that contribute dominantly to CSD (Eq. (9)) and their spectral density (Eq. (11)). We evaluated the weights of different modes normalized to zero mode $\beta_j/\beta_0$ and introduced a threshold value of 1%. The values of these mode weights are presented in Fig. 4 as a function of the mode number for all photon energies and energy spread values considered in this work. As it is seen from this figure for 500 eV photon energy and zero energy spread only three modes contribute significantly and this value is increased only to four modes as energy spread values are increased. Already for 12 keV photon energy, the number of modes with the contribution higher than 1% is about 10 modes. It is rising up with the increased energy spread values reaching 42 modes at $2 \cdot 10^{-3}$ relative energy spread value. The number of modes that contribute dominantly to CSD and spectral density with a threshold of 1% is 19 for 24 keV and 35 for 50 keV in zero relative energy spread case. With the increase of energy spread to the value of $2 \cdot 10^{-3}$ the number of modes contributing dominantly to CSD also increasing to 54 at 24keV and to 90 at 50 keV, which is significantly larger than in the previous case of lower photon energies.

As soon as all mode weights were determined by the XRT software we also determined the global degree of coherence for the considered photon energies and energy spread values according to Eq. (12) (see Table 4). The values of the global degree of coherence vary from 90% to 11% for photon energies from 500 eV to 50 keV, respectively. They drop down by 18% - 73% with the increase of energy spread from zero to $2 \cdot 10^{-3}$ relative energy spread value for the same range of photon energies.

### c) Coherent fraction of radiation

As soon as the mode values are normalized by the sum of all modes $\sum_{j=0}^{\infty} \beta_j(\omega)$ (as it is implemented in the XRT software), the value of the first mode naturally gives the coherent fraction of radiation (see Eq. (13)). The values of the coherent fraction are presented in Fig. 5 for all photon energies considered in this work (shown by triangles) as a function of electron beam emittance for different values of energy spread. We can see from this figure (see also Table 4) that at 10 pmrad we are getting very high coherence values about 95% at 500 eV and 55% at 12 keV at zero energy spread. With the increase of the energy spread these values become slightly lower at the energy spread values of $1 \cdot 10^{-3}$ (91% and 41%) and significantly lower at the energy spread values of $2 \cdot 10^{-3}$ (85% and 28%). At higher photon energies coherent fraction of the radiation drops from 40% to 26% at 24 keV and from 26% to 15% at 50 keV, while energy spread increases from zero to $2 \cdot 10^{-3}$.

We compared these results of XRT simulations with calculations of coherent fraction in the frame of the analytical approach (shown by circles in Fig. 5). Analytical simulations were performed by taking field amplitudes at the source and in the far-field region according to Eqs. (32-33) and performing mode decomposition similar to the XRT code (see for details the supporting information). We see from Fig. 5 that the results of the analytical approach fit well to the results of XRT simulations for all energies and energy spread values considered in this work.

### d) *Cross-spectral density function analyzed in one transverse direction*

Finally, we would like to compare the results obtained in the previous sections with the results of simulations of correlation functions in one direction. Coherent-mode representation of correlation functions, being very general, is providing an excellent theoretical insight into the problem. At the same time, the shape of the coherent modes is not easy to determine experimentally. Since the cross-spectral density $W(\boldsymbol{r_1}, \boldsymbol{r_2})$ is a 4D function, its determination from the experimental data is not a simple task. In most of the experiments (as Young's double pinhole experiment) correlation functions are determined in each direction separately (see, for example, Vartanyants et al., 2011; Singer et al., 2012, Skopintsev et al., 2014). Next, it is assumed that the CSD of the whole field may be represented as a product of its orthogonal directions (see Eq. (17)). We analysed this question for the case of the diffraction-limited source.

Simulations of the correlation functions in one transverse direction were performed in the far-field region at 30 m distance from the source. To determine these functions the XRT software and analytical expression of the wave field from a single electron in the far-field region (see Eq. (32)) were used. The results of these simulations for two photon energies of 500 eV and 12 keV are presented in Table 5 and Fig. 6 (results of the analytical approach for the same energies are shown in the supporting information Figs. S5-S6). The intensity distribution $I(x,y)$ (Fig. 6(a,b)), absolute value of the CSD in the horizontal direction $|W(x_1,x_2)|$ (Fig. 6(c,d)), absolute value of the SDC $|\mu(x_1,x_2)|$ (Fig. 6(e,f)), and absolute value of the spectral degree of coherence as a function of spatial separation of two points $|\mu(\Delta x)|$ (Fig. 6(g,h)) for both energies are presented in Fig. 6. It is well seen from this figure that the functional dependence of these parameters is non-Gaussian for the 500 eV photon energy. But already at 12 keV these parameters can be successfully described by Gaussian functions (the same is valid for higher energies). As a general rule, the more modes contribute to the CSD, the more this dependence resembles Gaussian. We determined the root mean square (rms) values $\sigma_{x,y}$ of the intensity distribution in the far-field region by Eqs. (34) (see Fig. 6 (a,b)). The values of the transverse degree of coherence $\zeta_x^{DC}$ (see Fig. 6 (c,d)) were calculated according to Eq. (5). Anti-diagonal cuts of the spectral degree of coherence function (see Fig. 6 (e,f)) were used to determine the coherence length of radiation $\xi_x$ as its rms values according to Eqs. (34). As it is seen from Fig. 6 results of simulations performed by the XRT software match extremely well to the ones performed analytically (see supporting information Figs. S5-S6).

As it is seen from our simulations, the CSD function $W(x_1,x_2)$ has a rectangular shape (see Fig. 6 (c,d)) because the coherence length of the beam is larger than the beam size at 500 eV. This is similar to our earlier observations in Ref. (Gorobtsov *et al.*, 2018). Also in this diffraction-limited case, the SDC shows strong oscillations at the tails of the beam profile due to the fact that the total photon radiation is defined mostly by characteristics of single-electron radiation. This is also a reason why Gaussian approximation is not valid in this diffraction-limited case and more careful analysis is required. Similar to our previous studies we also observed a decrease of the values of the transverse degree of coherence with the increase of energy spread. (see Table 5). Similar simulations were performed for higher photon energies of 24 keV, 50 keV and are summarized in the supporting information Figs. S3-S4 and in Table 5.

Note, though Gaussian approximation fits nicely for the spectral degree of coherence profile as well as for cross spectral density at higher energies, the global degree of coherence is not equal to the product of transverse coherence values $\zeta \neq \zeta_x \zeta_y$ (compare Tables 4 and 5). This leads to a conclusion that correlation functions for synchrotron radiation close to the diffraction limit are not separable in the two transverse directions.

## 5. Conclusions and outlook

In summary, we have provided a detailed analysis of the coherence properties of a high-energy synchrotron storage ring with ultra-low emittance values near 10 pmrad and a wide range of photon energies from 500 eV to 50 keV. Such low values of electron beam emittance are expected to be reached at PETRA IV facility (Schroer *et al.*, 2018; Schroer *et al.*, 2019). In addition, we analyzed the effect of electron energy spread on radiation properties of a low-emittance synchrotron ring for the same energy range. All simulations were performed by XRT software and were additionally compared with the results of analytical simulations based on equations of synchrotron radiation and also with approximated formulas given in Ref. (Tanaka & Kitamura, 2009). We note that all three approaches produced similar results for the whole set of parameters investigated in this work.

The following important lessons were learned during this work. In order to determine the properties of radiation from diffraction-limited sources, an approach based on statistical optics (Goodman, 1985; Mandel & Wolf, 1995) should be used. For low-emittance storage rings radiation field cannot be any more approximated by Gaussian functions. The single electron radiation distribution defines the beam profile in this low-emittance regime. As a consequence, a minimum photon emittance (diffraction limit) that may be reached on such storage rings is rather $\lambda/2\pi$ than $\lambda/4\pi$, typical for Gaussian beams. As a result, the degree of coherence of the radiation goes to its asymptotic limit and approaches unity when photon emittance is reaching the value of $\lambda/2\pi$.

Another lesson is that even for such low emittance values as 10 pmrad true diffraction limit will be reached, in fact, only at soft X-ray energies of about 500 eV. In this case, only a few modes will contribute to the radiation field, but already at 12 keV radiation field will consist of about ten modes and will start to be more Gaussian-like. This effect will be even more pronounced at high energies. In order to reach the true diffraction limit for hard X-rays emittance should be pushed down to about 1 pmrad, that is, unfortunately, out of reach for present technology.

Our results also show that correlation functions, describing radiation field properties such as the degree of coherence, cannot be factorized into two transverse directions for these low-emittance sources. For a full description of the radiation properties of these sources eigenvalue decomposition of the radiation field has to be performed which offers good theoretical insight as well as complete generality. It also means that present experimental approaches which measure coherence properties of radiation in each direction separately (Vartanyants et al., 2011; Singer et al., 2012; Skopintsev et al., 2014) should be generalized to 2D methods of coherence determination.

Another outcome of our work is the analysis of the electron energy spread effects on coherence properties of low-emittance storage rings. We demonstrate that this effect becomes more noticeable for low electron beam emittance. The larger the energy spread values, the more the source size and divergence are affected, and, as a consequence, the degree of coherence and coherent fraction value of the radiation are decreased. We found that, in order to keep high coherence values of radiation, the relative energy spread should not exceed the value of $1 \cdot 10^{-3}$.

Finally, our results demonstrate that the coherence properties of the future diffraction-limited sources will be outstanding (Hettel, 2014; Weckert, 2015). We hope that the general approach and new tools for an adequate description of the coherence properties of synchrotron sources, provided in this work, will be helpful for the design and planning of future diffraction-limited sources worldwide.

**Acknowledgements**

The authors would like to thank E. Weckert for fruitful discussions and support of the project, R. Chernikov and K. Klementiev for the support with the XRT software and discussion of the results, M. Tischer, G. Geloni and Yu. Obukhov for careful reading of the manuscript, E. Saldin for fruitful discussions, RKh and IAV acknowledge the support by the Helmholtz Associations Initiative and Networking Fund and the Russian Science Foundation (Project No. 18-41-06001).

**Table 1.**

Basic simulation parameters of accelerator and undulator source for all photon energies.

| | |
|---|---|
| Electron Energy | 6.0 GeV |
| Beam current | 100 mA |
| Horizontal and vertical electron beam emittance | 10 pmrad |
| Horizontal and vertical betatron functions, $\beta_x$, $\beta_y$ | 2.0 m |
| Relative energy spread values | 0; $1 \cdot 10^{-3}$; $2 \cdot 10^{-3}$ |
| Undulator length | 5 m |

**Table 2.**

Phase space parameters (in Gaussian approximation) of an X-ray source at different photon energies for 10 pmrad electron beam emittance and zero energy spread value.

| Photon energy, keV | **0.5** | **12** | **24** | **50** |
|---|---|---|---|---|
| Number of undulator periods, $N_u$ | 72 | 170 | 170 | 170 |
| Number of harmonics, $n$ | $1^{st}$ | $3^{rd}$ | $3^{rd}$ | $5^{th}$ |
| Single electron radiation size $\sigma_r$ , $\mu m$, [Eq.27] | 12.5 | 2.55 | 1.79 | 1.25 |
| Single electron radiation divergence $\sigma'_r$ , $\mu rad$, [Eq. (27)] | 15.7 | 3.2 | 2.3 | 1.5 |
| Single electron radiation emittance $\varepsilon_r = \lambda/4\pi$ , pmrad [Eq. (29)] | 197 | 8.2 | 4.1 | 1.9 |
| Electron beam size $\sigma_e$ , $\mu m$ | 4.47 | 4.47 | 4.47 | 4.47 |
| Electron beam divergence $\sigma'_e$ , $\mu rad$ | 2.24 | 2.24 | 2.24 | 2.24 |
| Total photon emittance $\varepsilon_{ph}$, pmrad [Eq. (26)] | 211 | 20 | 15.5 | 12.5 |

**Table 3.**

Photon emittance values determined at different energies and zero electron energy spread using XRT simulations, analytical approach, and Eqs. (30-31). Results are presented for the 10 pmrad and 1 pmrad electron emittance values.

| Photon energy, keV | **0.5** | **12** | **24** | **50** |
|---|---|---|---|---|
| $\varepsilon_{coh} = \lambda/2\pi$, pmrad | 395 | 16.4 | 8.2 | 3.9 |
| *10 pmrad natural electron emittance* | | | | |
| $\varepsilon_{ph}{}^{XRT}$, pmrad | 398 | 24.6 | 18.5 | 15.2 |
| $\varepsilon_{ph}{}^{Analyt.}$, pmrad | 402 | 24.9 | 17.7 | 13.8 |
| $\varepsilon_{ph}{}^{T\&K}$, pmrad | 403 | 26.5 | 18.3 | 13.8 |
| Source size (FWHM) $\Sigma_{ph}^{Analyt.}$, μm | 59 | 15.9 | 13.5 | 12 |
| Source divergence (FWHM) $\Sigma'_{ph}{}^{Analyt.}$, μm | 37 | 9.3 | 7.5 | 6.5 |
| *1 pmrad natural electron emittance* | | | | |
| $\varepsilon_{ph}{}^{XRT}$, pmrad | 380 | 20 | 10 | 5 |
| $\varepsilon_{ph}{}^{Analytical}$, pmrad | 391 | 16.3 | 8.9 | 5 |
| $\varepsilon_{ph}{}^{T\&K}$, pmrad | 396 | 17.5 | 9.3 | 5 |
| Source size (FWHM) $\Sigma_{ph}^{Analyt.}$, μm | 58.6 | 12.4 | 9.1 | 6.7 |
| Source divergence (FWHM) $\Sigma'_{ph}{}^{Analyt.}$, μm | 37.5 | 7.8 | 5.7 | 4.1 |

**Table 4.**

Global degree of coherence (Eq. (12)) and the coherent fraction (Eq. (13)) determined from coherent mode decomposition using XRT software and analytical approach of Eq. (32). All simulations are performed at the electron beam emittance value of 10 pmrad for different photon energies and energy spread values.

| Photon energy, keV | 0.5 | 12 | 24 | 50 |
|---|---|---|---|---|
| Zero relative energy spread | | | | |
| Global degree of coherence $\zeta_{XRT}^{DC}$ | 0.90 | 0.34 | 0.20 | 0.11 |
| Coherent fraction $\zeta_{XRT}^{CF}$ | 0.95 | 0.55 | 0.40 | 0.26 |
| Coherent fraction $\zeta_{Analyt.}^{CF}$ | 0.95 | 0.56 | 0.41 | 0.26 |
| $1 \cdot 10^{-3}$ relative energy spread | | | | |
| Global degree of coherence $\zeta_{XRT}^{DC}$ | 0.84 | 0.20 | 0.13 | 0.06 |
| Coherent fraction $\zeta_{XRT}^{CF}$ | 0.91 | 0.41 | 0.35 | 0.22 |
| Coherent fraction $\zeta_{Analyt.}^{CF}$ | 0.93 | 0.37 | 0.3 | 0.17 |
| $2 \cdot 10^{-3}$ relative energy spread | | | | |
| Global degree of coherence $\zeta_{XRT}^{DC}$ | 0.74 | 0.11 | 0.07 | 0.03 |
| Coherent fraction $\zeta_{XRT}^{CF}$ | 0.85 | 0.28 | 0.26 | 0.15 |
| Coherent fraction $\zeta_{Analyt.}^{CF}$ | 0.89 | 0.26 | 0.21 | 0.12 |

**Table 5.**

Degree of coherence in one transverse direction obtained from the XRT simulations at 10 pmrad electron beam emittance compared with the analytical analysis for the different photon energies and relative energy spread values.

| Photon energy, keV | 0.5 | 12 | 24 | 50 |
|---|---|---|---|---|
| Zero relative energy spread | | | | |
| Degree of coherence $\zeta_x^{XRT}$ | 0.92 | 0.37 | 0.26 | 0.15 |
| Degree of coherence $\zeta_x^{Analyt.}$ | 0.92 | 0.39 | 0.25 | 0.15 |
| Degree of coherence $\zeta_x^{GSM}$ | 0.87 | 0.39 | 0.28 | 0.16 |
| $1 \cdot 10^{-3}$ relative energy spread | | | | |
| Degree of coherence $\zeta_x^{XRT}$ | 0.89 | 0.25 | 0.17 | 0.11 |
| Degree of coherence $\zeta_x^{Analyt.}$ | 0.89 | 0.23 | 0.17 | 0.09 |
| Degree of coherence $\zeta_x^{GSM}$ | 0.85 | 0.29 | 0.19 | 0.11 |
| $2 \cdot 10^{-3}$ relative energy spread | | | | |
| Degree of coherence $\zeta_x^{XRT}$ | 0.82 | 0.17 | 0.12 | 0.07 |
| Degree of coherence $\zeta_x^{Analyt.}$ | 0.80 | 0.15 | 0.12 | 0.07 |
| Degree of coherence $\zeta_x^{GSM}$ | 0.84 | 0.21 | 0.14 | 0.09 |

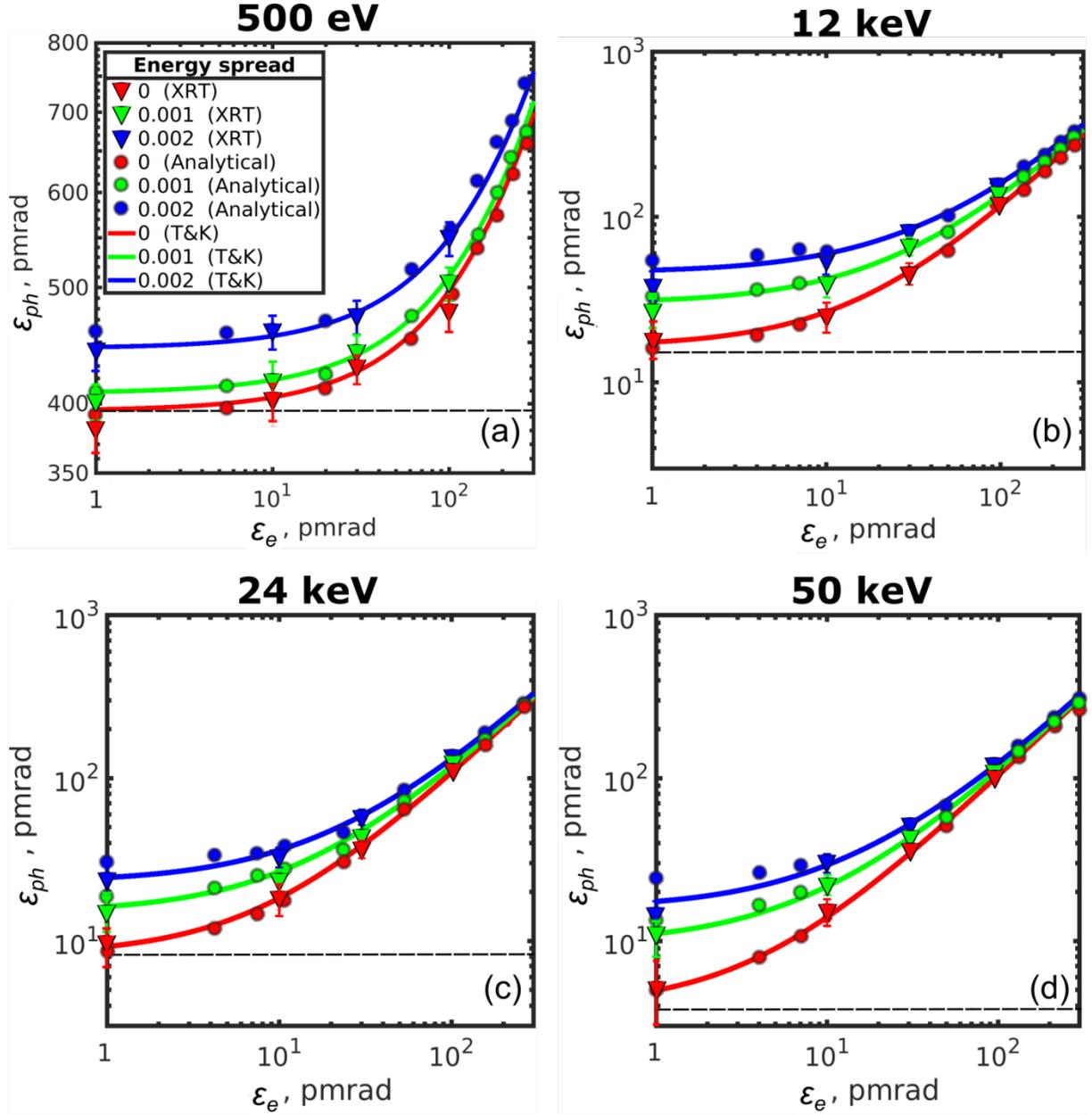

**Figure 1**. Photon emittance $\varepsilon_{ph}$ as a function of the electron beam emittance $\varepsilon_e$ for the different values of the photon energy and energy spread in one transverse direction. Triangles are XRT simulations, circles are analytical calculations and lines are the values obtained from the Tanaka and Kitamura (T&K) approach (Eqs. (30-31)). Red, green, and blue color correspond to 0, $1\cdot10^{-3}$, and $2\cdot10^{-3}$ relative energy spread values, respectively. Note the different scale for 500 eV emittance value. Dashed horizontal line corresponds to the value of the photon emittance of $\lambda/2\pi$.

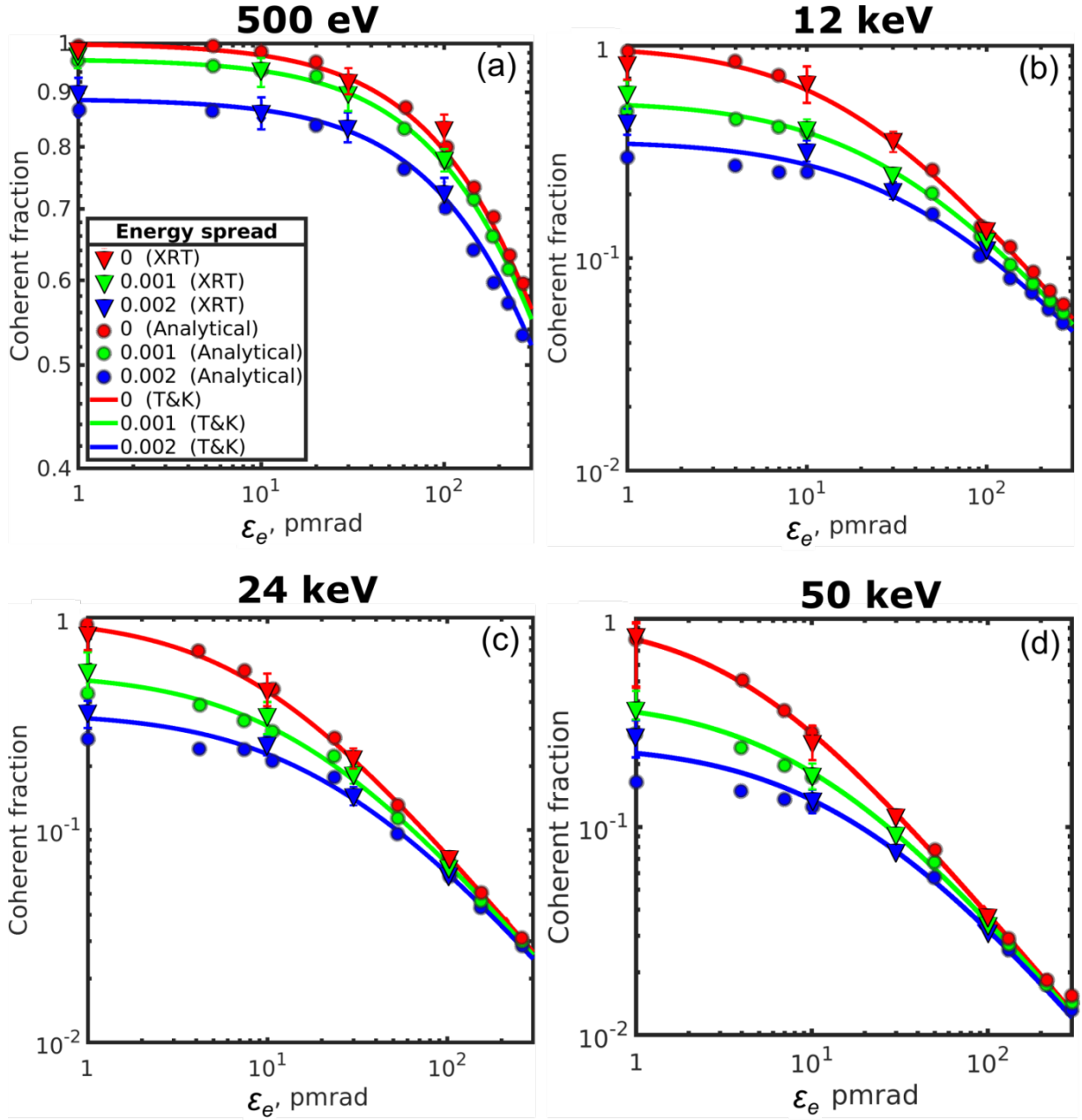

**Figure 2**. Coherent fraction of radiation $\zeta^{CF}$ as a function of the electron beam emittance $\varepsilon_e$ for the different values of the photon energy and energy spread in one transverse direction calculated according to Eq. (37). Triangles are XRT simulations, circles are analytical calculations, and lines are the values obtained the Tanaka and Kitamura (T&K) approach (Eqs. (30-31)). Red, green, and blue color correspond to 0, $1\cdot10^{-3}$, and $2\cdot10^{-3}$ relative energy spread values, respectively. Note the different scale for 500 eV coherent fraction value.

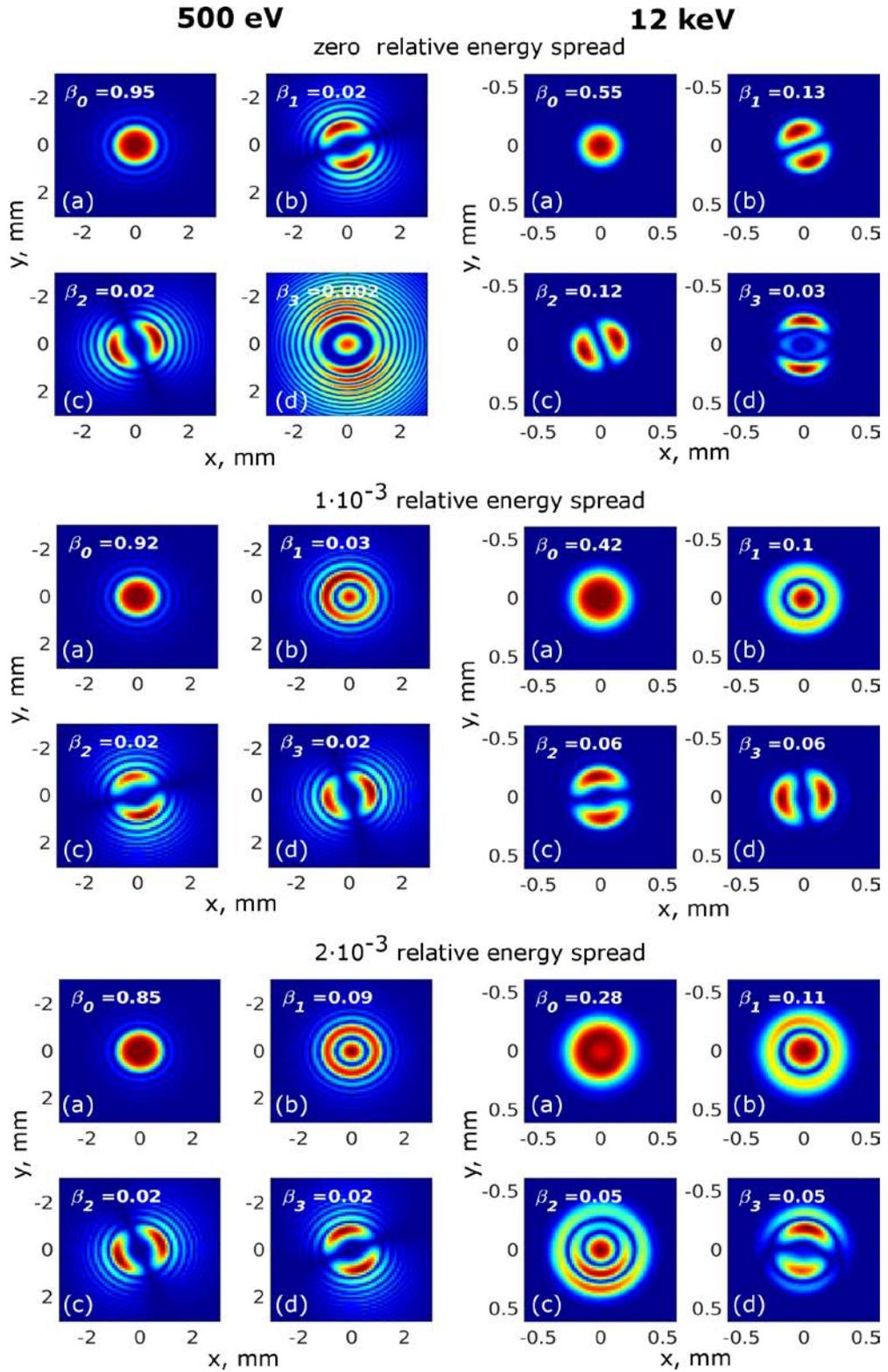

**Figure 3**. First four modes and their normalized weights $\beta_j$ obtained from the coherent mode decomposition of the CSD at 500 eV (left column) and 12 keV (right column) photon energy for three different relative energy spread values obtained by XRT simulations.

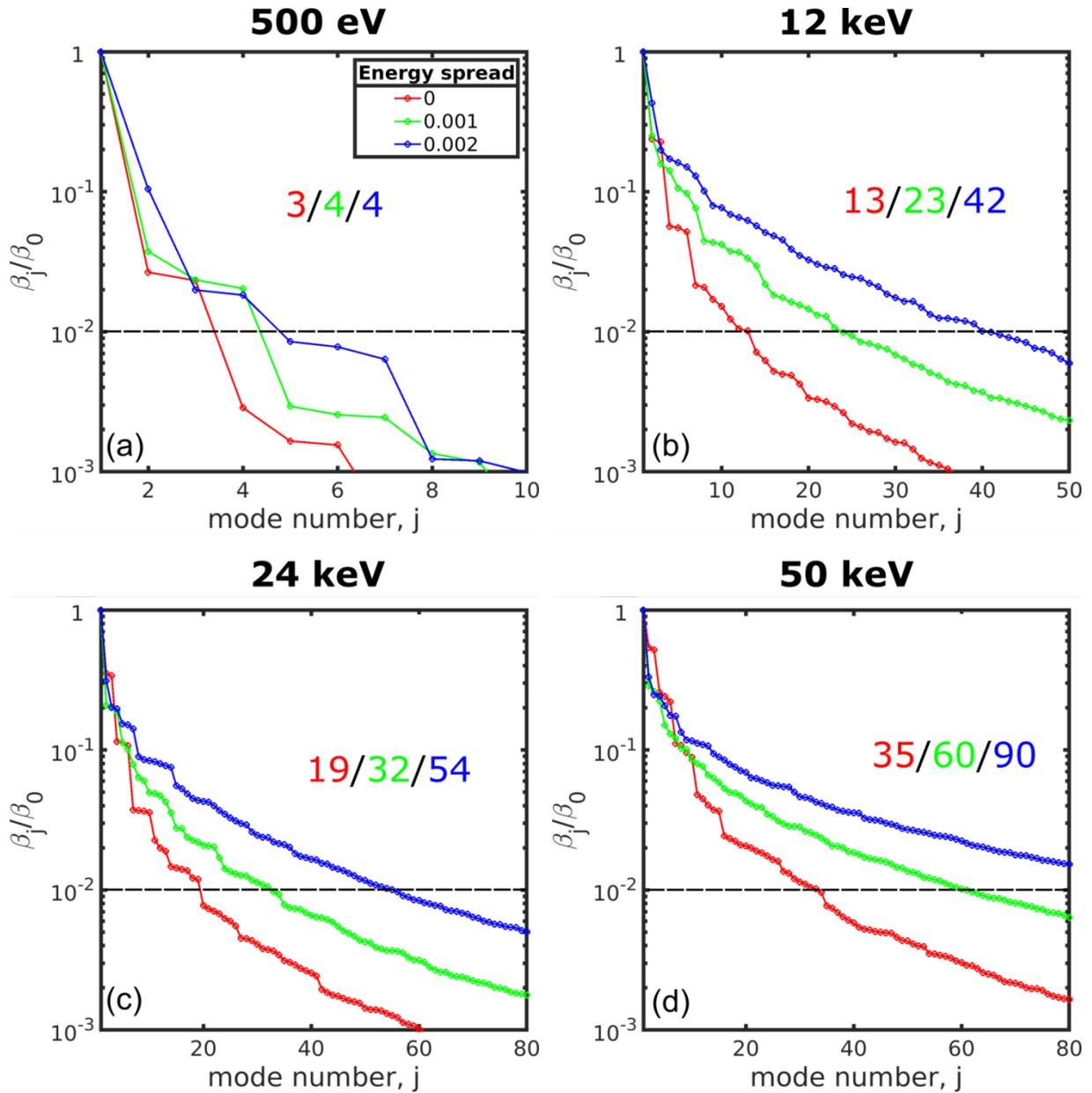

**Figure 4**. Weights of different modes normalized to the weight of a zero-mode $\beta_j/\beta_0$ as a function of the mode number $j$ for the different values of the photon energy and energy spread. Horizontal dashed line corresponds to the value of 1%. Points are connected by lines for better visibility. Red, green, and blue color correspond to 0, $1\cdot10^{-3}$, and $2\cdot10^{-3}$ relative energy spread values, respectively. Number of modes exceeding 1% threshold is given in each panel.

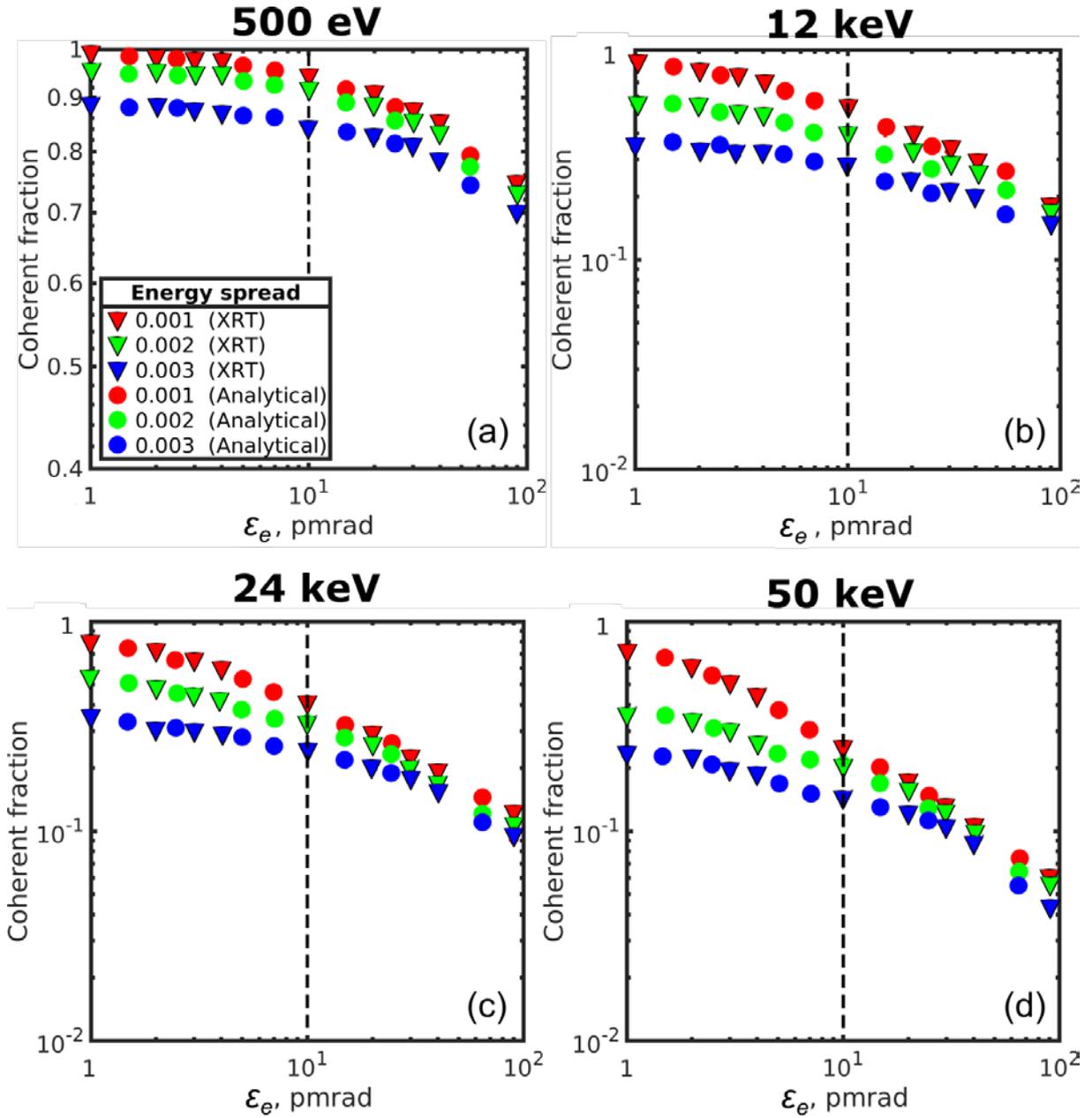

**Figure 5**. Coherent fraction of radiation $\zeta^{CF}$ as a function of the electron beam emittance $\varepsilon_e$ for the different values of the photon energy and energy spread in one transverse direction. Triangles are XRT simulations, circles are analytical calculations performed according to Eq. (13). Red, green, and blue color correspond to 0, $1\cdot10^{-3}$, and $2\cdot10^{-3}$ relative energy spread values, respectively. Dashed vertical line corresponds to the value of the electron emittance of 10 pmrad. Note the different scale for 500 eV coherent fraction value.

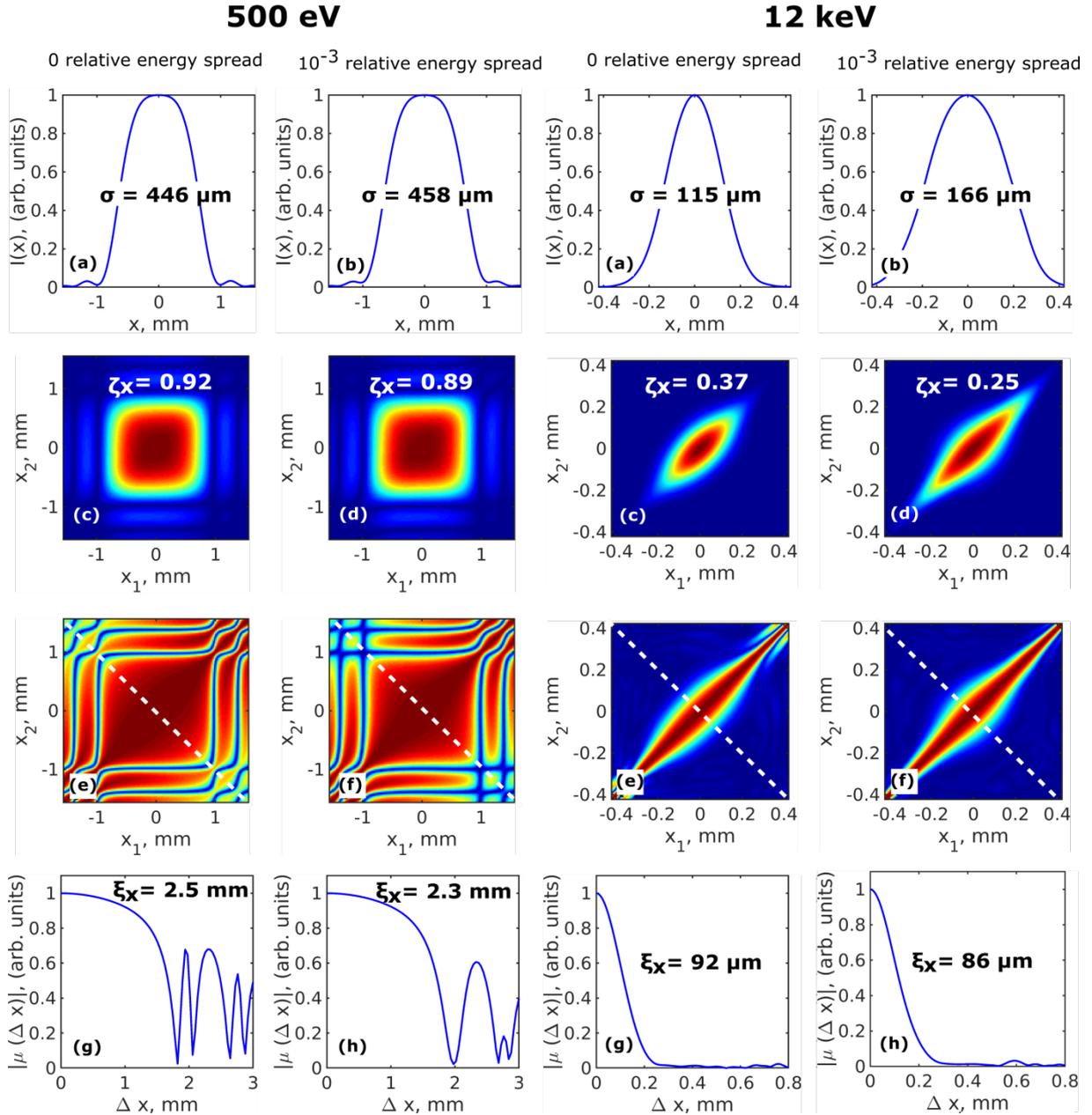

**Figure 6.** Simulations of the correlation functions in horizontal direction performed by the XRT software for 500 eV (left column) and 12 keV (right column) photon energy. Intensity distribution I(x) (a,b), absolute value of the cross-spectral density in the horizontal direction |W($x_1,x_2$)| (c,d), absolute value of the SDC |μ($x_1,x_2$)| (e,f), and absolute value of the spectral degree of coherence along the anti-diagonal line (shown in (e,f)) as a function of separation of two points |μ(Δx)| (g,h) simulated in horizontal direction 30 m downstream from the undulator source. In (a,b) σ is the rms value of the beam size, in (c,d) $ζ_x$ is the transverse degree of coherence, in (g,h) $ξ_x$ is the coherence length determined in horizontal direction.

# Supporting information for:

# Coherence properties of the high-energy fourth-generation X-ray synchrotron sources

R. Khubbutdinov[1,2], A. P. Menushenkov[2], and I. A. Vartanyants [1,2,*]

[1]*Deutsches Elektronen-Synchrotron DESY, Notkestrasse 85, D-22607 Hamburg, Germany*

[2]*National Research Nuclear University MEPhI (Moscow Engineering Physics Institute)*

*Kashirskoe shosse 31, 115409 Moscow, Russia*

July 8, 2019

### S1. Details of XRT simulations

*a) Field amplitude and source parameters simulation*

To fully characterize source properties one needs to know the amplitude and intensity distributions at the source position and in the far field region, which were simulated at first by means of XRT software. The initial step was to determine synchrotron facility parameters as well as parameters of the undulator according to Table 1. The XRT software calculates an amplitude of the radiation for a single electron from the undulator source at the given photon energy (Jackson, 1962) on a given angular mesh (see Table S1). The transverse field from each electron gets individual random angular and coordinate offsets within the emittance distribution assuming Gaussian statistics of the electron bunch. An individual random shift to gamma Lorentz factor within the energy spread is counted as well (Klementiev & Chernikov, 2014).

Amplitude and intensity simulations for all photon energies considered in this work were performed in the far-field at a distance of 30 m from the source which is corresponding to Fresnel numbers given in Table S1. For a defined number of electrons (see Table S1) their corresponding amplitudes were stored in the matrix for further analysis of coherence functions. The resulting intensity was determined as a sum of intensities of each electron.

* corresponding author: Ivan.Vartanyants@desy.de

Angular divergence of the source was determined as the variance value of the intensity distribution at the far field position (Eq. (34)). Transverse positional distribution at the source position was obtained by taking the Fourier transform of the angular field distribution in the far field. The size of the source was determined as the variance value of the intensity distribution at the source position in the middle of the undulator (Eq. (34)). Total photon emittance was calculated as the product of the source size and source divergence (Eq. 28) (see Fig. 1 (triangles)). The error of the performed simulation with the XRT software for each natural electron emittance value was calculated as the standard deviation of the corresponding value.

### b) Coherent-mode representation

To determine independent coherent modes of radiation or perform coherent mode decomposition one needs to solve Fredholm integral equation according to Eq. (10). We performed mode decomposition of the CSD according to Eqs. (9-11) using the XRT software. In the XRT software, CSD is calculated according to Eqs. (7, 8) for the radiation in the far field region for a single electron. The CSD of radiation is, in general, a four-dimensional (4D) matrix (the CSD function depends on coordinates $x_1$, $y_1$; $x_2$, $y_2$). Effectively in the XRT software, one more dimension is added by including a number of electrons so that the correlation function is saved for each electron, which gives additional dimension. The next step is to compose the 4D CSD matrix from the 5D field distribution and to solve the CSD for eigenvalues and eigenfunctions.

The eigenvalue problem for the CSD matrix in XRT software is reduced to two-dimensional (2D) by applying principal component analysis (PCA) (see Fig. S1). After the diagonalization of this matrix, the eigenvalues (mode weights) and eigenfunctions (modes) of CSD were determined (see Fig. 3). If the mode weights are known than the global degree of coherence and the coherent fraction of radiation may be calculated according to Eqs. (12) and (13) (see Fig. 5 (triangles)). It should be also noted, that the global degree of coherence according to the PCA method can be obtained without solving this huge eigenvalue problem. The global degree of coherence, in this case, can be calculated as the ratio between traces of the CSD matrix $\zeta^{DC} = \mathrm{Tr}(W_{PCA}^2) / \left( \mathrm{Tr}(W_{PCA}) \right)^2$, which is equivalent to Eq. (12), where $W_{PCA}$ is the rearranged CSD matrix according to the PCA method.

The number of electrons used in the coherent-mode simulations is given in Table S1. The first four coherent modes and their normalized weights for high photon energies of 24 keV and 50 keV for different energy spread values are shown in Fig. S2. The behavior of these modes is similar to the ones observed at 500 eV and 12 keV case (see Fig. 3). The values of the global degree of coherence are given in Table 4.

*c) Degree of transverse coherence*

Simulations of amplitudes and correlation functions in one transverse direction were performed by XRT software as well. One dimensional CSD function at given photon energy was simulated in the far field region as average over an ensemble of electrons according to Eq. (8) in the horizontal direction. Parameters of the simulation are given in Table 5. Results of simulations for high photon energies of 24 keV and 50 keV are shown in Fig. S3 and Fig. S4. As it is seen from these figures, energy spread enlarges rms values $\sigma$ of the intensity distribution (calculated as the variance value, according to Eq. (34)) (Fig. S3-S4 (a,b,c)), while it narrows the CSD and the spectral degree of coherence in anti-diagonal direction (Fig. S3-S4(d-i)), which leads to a decrease of coherence. The values of the degree of transverse coherence in these cases are given in Table 5.

**S2. Details of analytical simulations**

*a) Field amplitude simulation*

In the frame of the analytical approach, source properties were calculated as well (Geloni *et al.*, 2015; Geloni, 2018). In this case, the amplitude $E(\theta_x, \theta_y)$ in the far field region from a single electron was calculated according to Eq. (33). Total amplitude in the far field region for the electron bunch was defined as

$$E_{total}(\theta_x, \theta_y) = \int E(\theta_x, \theta_y) f_{\eta_x}(\eta_x) f_{\eta_y}(\eta_y) f_{l_x}(l_x) f_{l_y}(l_y) f_{\gamma_E}(\gamma_E) d\eta_x d\eta_y dl_x dl_y d\gamma_E ,$$

(1)

with

$$f_{\eta_x}(\eta_x) = \frac{1}{\sigma'_e \sqrt{2\pi}} e^{-\frac{\eta_x^2}{2\sigma_e'^2}} , \qquad f_{\eta_y}(\eta_y) = \frac{1}{\sigma'_e \sqrt{2\pi}} e^{-\frac{\eta_y^2}{2\sigma_e'^2}} ,$$

(2)

$$f_{l_x}(l_x) = \frac{1}{\sigma_e\sqrt{2\pi}}e^{-\frac{l_x^2}{2\sigma_e^2}}, \qquad f_{l_y}(l_y) = \frac{1}{\sigma_e\sqrt{2\pi}}e^{-\frac{l_y^2}{2\sigma_e^2}},$$

(3)

$$f_{\gamma_E}(\gamma_E) = \frac{1}{\sigma_\gamma\sqrt{2\pi}}e^{-\frac{\gamma_E^2}{2\sigma_\gamma^2}},$$

(4)

where $f_{\eta_x}(\eta_x)$, $f_{\eta_y}(\eta_y)$ are the angular offset distributions, $f_{l_x}(l_x)$, $f_{l_y}(l_y)$ are the coordinate offset distributions from the undulator axis and $f_{\gamma_E}(\gamma_E)$ is the electron energy distribution according to given energy spread. Effectively it means that each electron got individual random angular and coordinate offsets within the emittance distribution, and random shift to gamma Lorentz factor within the energy spread similar to XRT simulations. Additionally, the 2D amplitude distribution was saved for each electron into 3D matrix, where the third dimension is connected to the number of electrons. The amplitude at the source position $E_0(x, y)$ (in the middle of an undulator) was obtained by the use of the propagator, according to Eq. (33) in the main text.

*b) Source parameters simulation*

Calculations of source parameters were done for zero, $1 \cdot 10^{-3}$, and $2 \cdot 10^{-3}$ relative energy spread values. Amplitude and intensity distribution in the far field region at a distance of 30 m from the source for different photon energies were simulated on the same angular mesh, as in the case of XRT simulations (see Table S1). Total photon emittance was calculated in the same way as it was done previously according to Eq. (34) and Eq. (28) of the main text (see Fig. 1 (dots)). Due to the narrowing of divergence of the radiation higher is the photon energy more electrons were needed to sample the radiation on the virtual detector in order to obtain uniform intensity distribution. In order to accumulate good statistics and get a uniform intensity distribution number of electrons used in this calculation was increased (see Table 5). This adjustment is also valid for larger energy spread values.

*c) Coherent-mode representation, coherent fraction and degree of transverse coherence.*

The same parameters of the analytical simulations of the CSD function were used for XRT simulations (see Table S1). Correlation functions were calculated and saved for each electron according to Eq. (8), with the amplitude distribution calculated according to Eq. (32). The eigenvalue problem for the 5D CSD matrix was also reduced to two-dimensional by applying PCA as it was done previously by XRT software. In this case, a coherent fraction of the radiation as a function of electron emittance is shown in Fig. 5 (dots). In order to accumulate good statistics the number of electrons used in this calculation was also increased (see Table S1).

Simulations of the amplitude and CSD were also performed in one transverse direction in the frame of the analytical approach (see Figs. S5-S8). One dimensional amplitude distributions for all photon energies considered in this work are shown in Figs. S5-S8 (a,b,c). One dimensional CSD was calculated in the far field region as average over an ensemble of electrons according to Eq. (8) of the main text in the horizontal direction (see Figs. S5-S8 (d,e,f)). The degree of transverse coherence was calculated according to Eq. (19) of the main text. As it is seen from these figures effect of energy spread again enlarges rms values $\sigma$ of the intensity distribution at the same time it narrows the spectral degree of coherence in anti-diagonal direction, which also leads to a decrease in coherence value as it was already shown by XRT simulations. The values of the degree of transverse coherence obtained in the frame of the analytical approach are listed in Table 4. These results are very close to the results obtained from simulations by XRT software.

Due to the fact that analytical analysis matches very well to the simulation of correlation functions performed by XRT software (see Fig. 6 and Figs. S5-S8), it has been extended to other natural electron emittances in a range from 1 to 100 pmrad. The degree of transverse coherence (Eq. (19) of the main text) calculated in the frame of the analytical approach as a function of natural electron emittance for various relative energy spread values is shown in Fig. S9. We observe the same tendencies for the degree of transverse coherence as for coherent fraction of the radiation shown above in Fig. 2 and Fig. 5. At 500 eV and 12 keV we reach diffraction-limited case already at 10 pmrad and 1 pmrad respectively. However, at 24keV and 50 keV even 1 pmrad electron emittance is not sufficient to reach diffraction limit.

**Table S1.**

Details of XRT simulations and analytical approach. $N_e$ is the number of electrons used in the simulations of photon emittance and simulations of CSD.

| Energy, keV | $N_e$ (photon emittance) | | $N_e$ (coherent-mode decomposition) | | Angular mesh (AxA) | | | Size of the virtual detector, mm | Energy spread, $10^{-3}$ | Fresnel number |
|---|---|---|---|---|---|---|---|---|---|---|
| | XRT | Analytical | XRT | Analytical | Photon Emittance | Mode decomposition | | | | |
| | | | | | XRT & Analytical | XRT | Analytical | | | |
| 0.5 | $5\cdot10^3$ | $5\cdot10^3$ | $5\cdot10^3$ | $10^5$ | 512 | 128 | 256 | 5.4 | 0 | 0.05 |
| | | | | | | | | | 1 | 0.05 |
| | | | | | | | | | 2 | 0.06 |
| 12 | $5\cdot10^3$ | $10^4$ | $5\cdot10^3$ | $10^5$ | 512 | 128 | 256 | 2.4 | 0 | 0.09 |
| | | | | | | | | | 1 | 0.14 |
| | | | | | | | | | 2 | 0.21 |
| 24 | $5\cdot10^3$ | $1.5\cdot10^4$ | $5\cdot10^3$ | $10^5$ | 512 | 128 | 256 | 1.8 | 0 | 0.13 |
| | | | | | | | | | 1 | 0.19 |
| | | | | | | | | | 2 | 0.26 |
| 50 | $5\cdot10^3$ | $1.5\cdot10^4$ | $5\cdot10^3$ | $10^5$ | 512 | 128 | 256 | 1.2 | 0 | 0.19 |
| | | | | | | | | | 1 | 0.3 |
| | | | | | | | | | 2 | 0.44 |

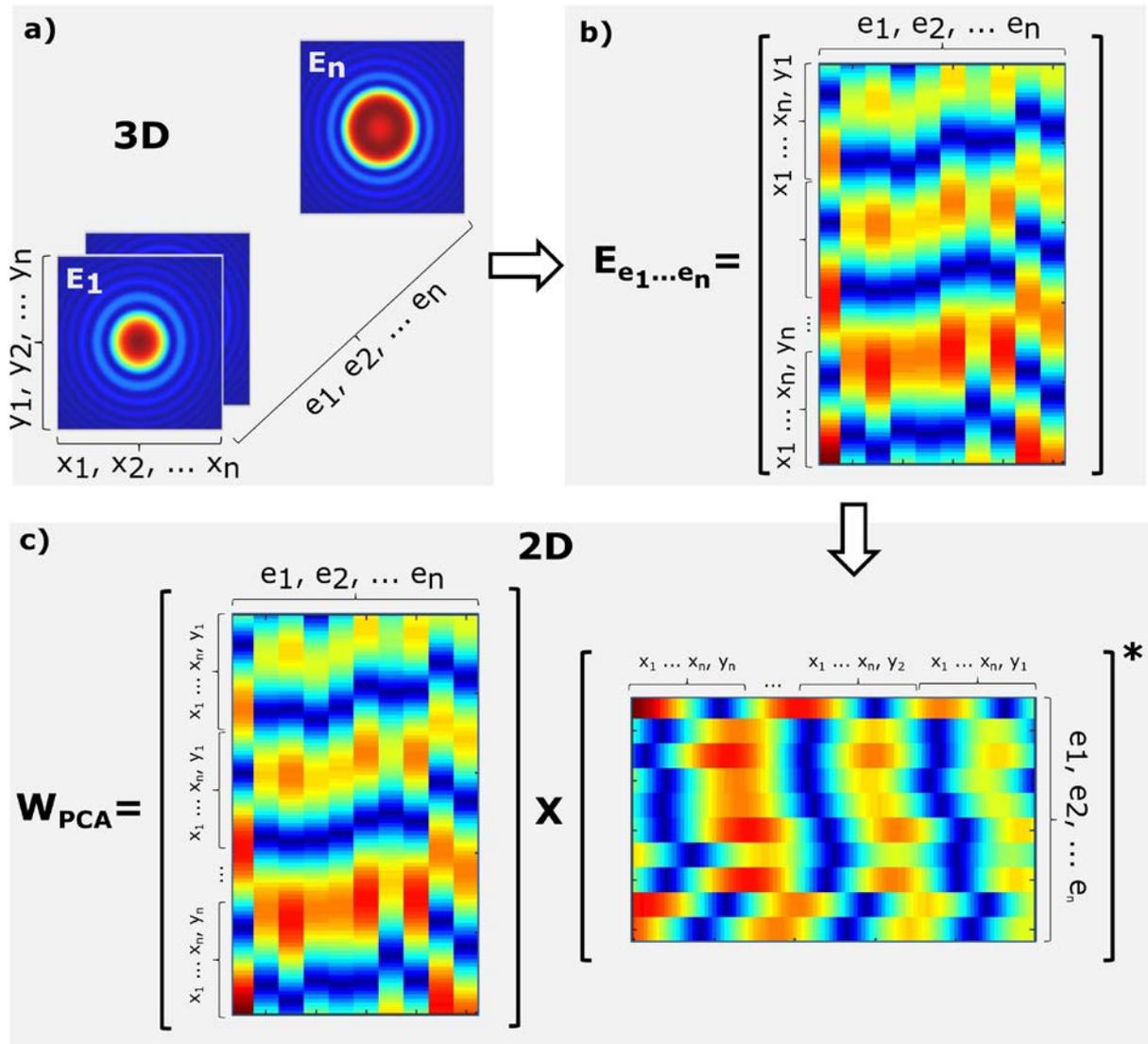

**Figure S1** PCA method used in the coherent mode decomposition. a) 2D amplitudes stored in a 3D matrix for each electron $e_1...e_n$, where $3^{rd}$ dimension is connected to the number of electrons. b) 3D matrix of amplitudes, rearranged to 2D matrix $E_{e_1-e_n}$, where each column contains all spatial information about an amplitude corresponding to one of the electrons $e_i$ from the electron bunch. c) Cross-spectral density matrix $W_{PCA}$ obtained by multiplication of matrix $E_{e_1-e_n}$ by its complex conjugated and transposed matrix.

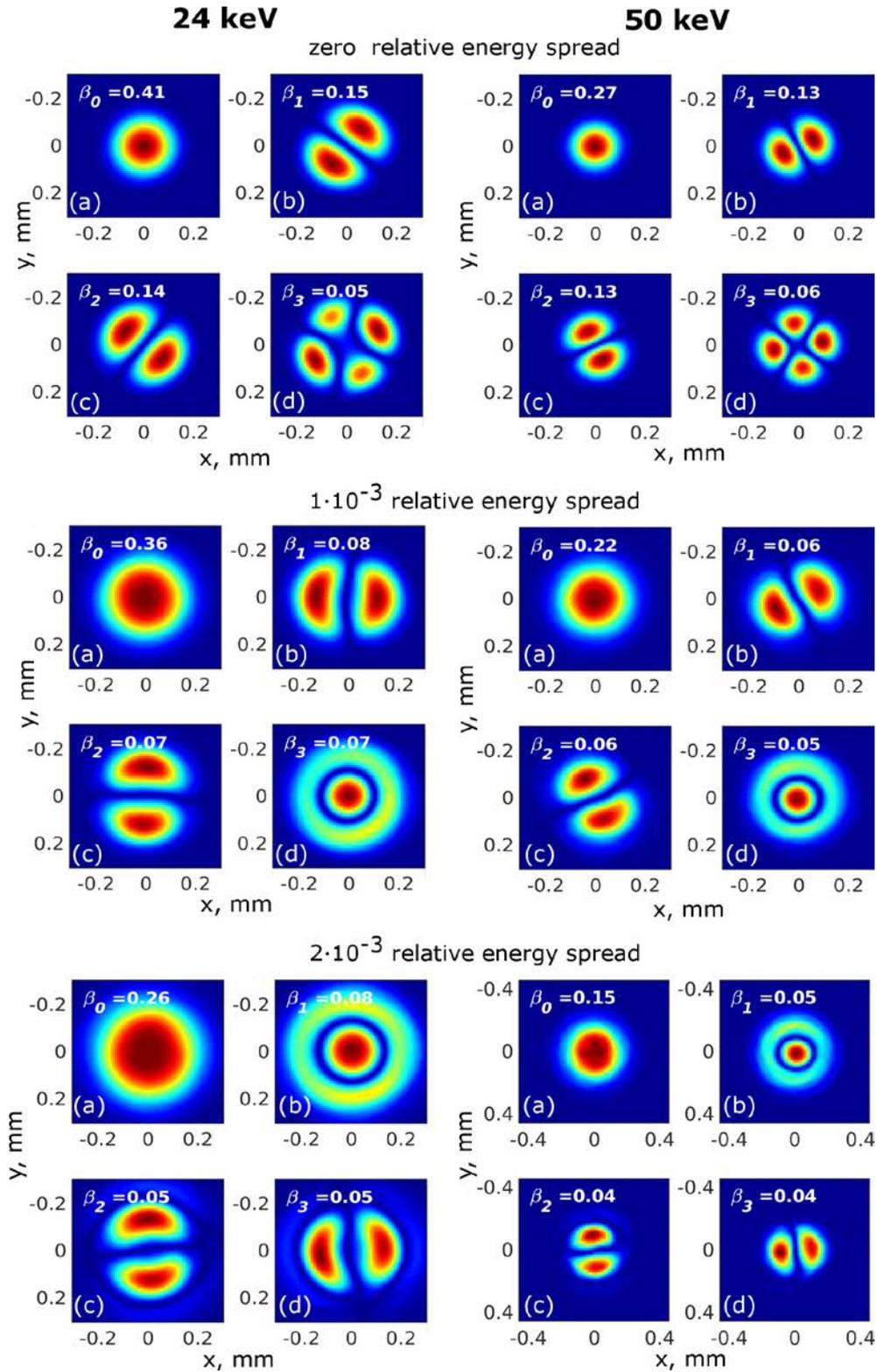

**Figure S2** First four modes and their normalized weights $\beta_j$ obtained from the coherent mode decomposition of the CSD at 24 keV (left column) and 50 keV (right column) photon energy for three different relative energy spread values obtained by XRT simulations.

## 24 keV

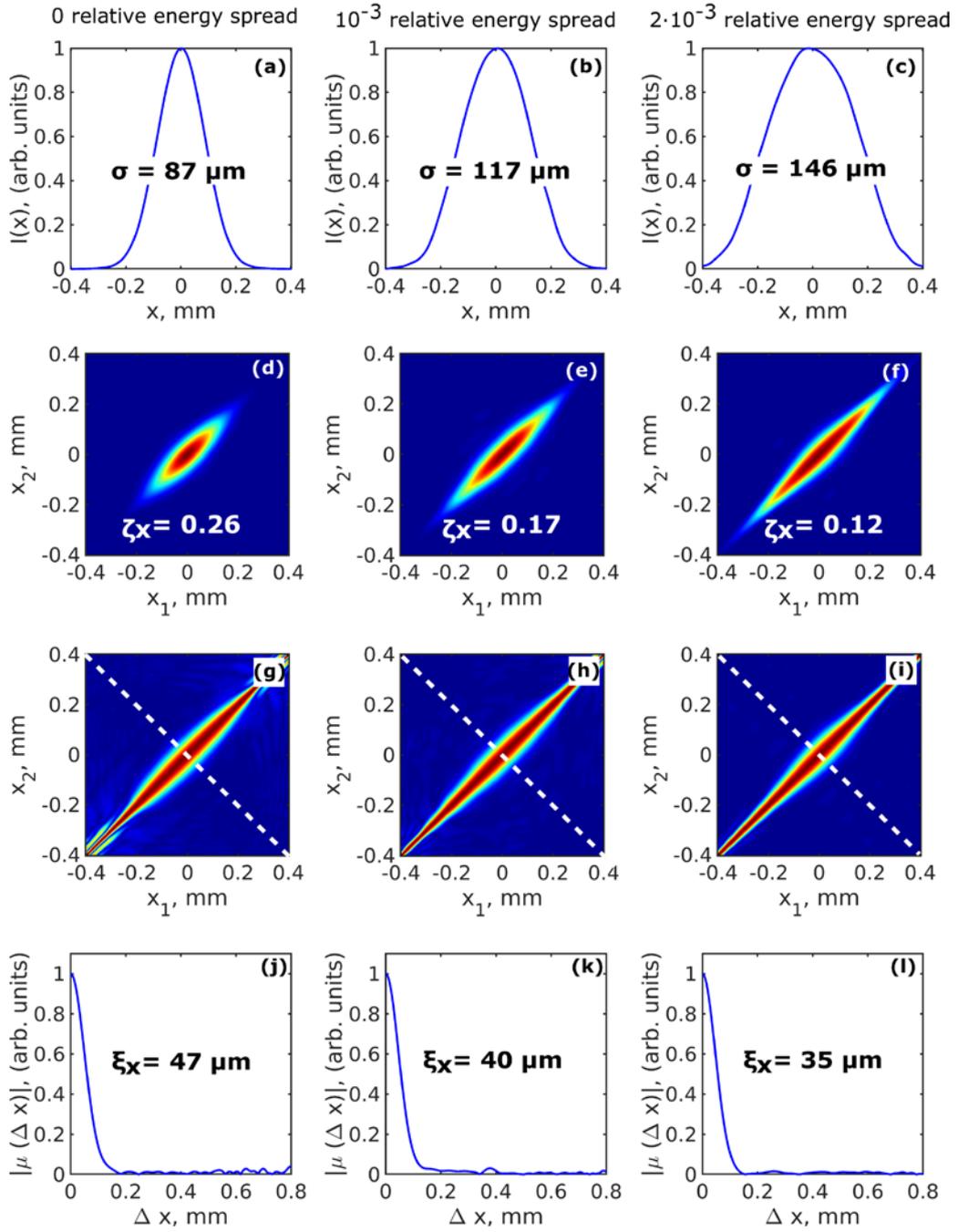

**Figure S3** Simulations of the correlation functions in horizontal direction performed by the XRT software for 24 keV photon energy and different energy spread values. Intensity distribution I($x$) (a-c), absolute value of the CSD in the horizontal direction |W($x_1,x_2$)| (d-f), absolute value of the SDC |μ($x_1,x_2$)| (g-i), and absolute value of the SDC along the anti-diagonal line (shown in (g-i)) as a function of separation of two points |μ($Δx$)| (j-l) simulated in horizontal direction 30 m downstream from the undulator source. In (a-c) $\sigma$ is the rms value of the beam size, in (d-f) $\zeta_x$ is the transverse degree of coherence, in (j-l) $\xi_x$ is the coherence length determined in the horizontal direction.

# 50 keV

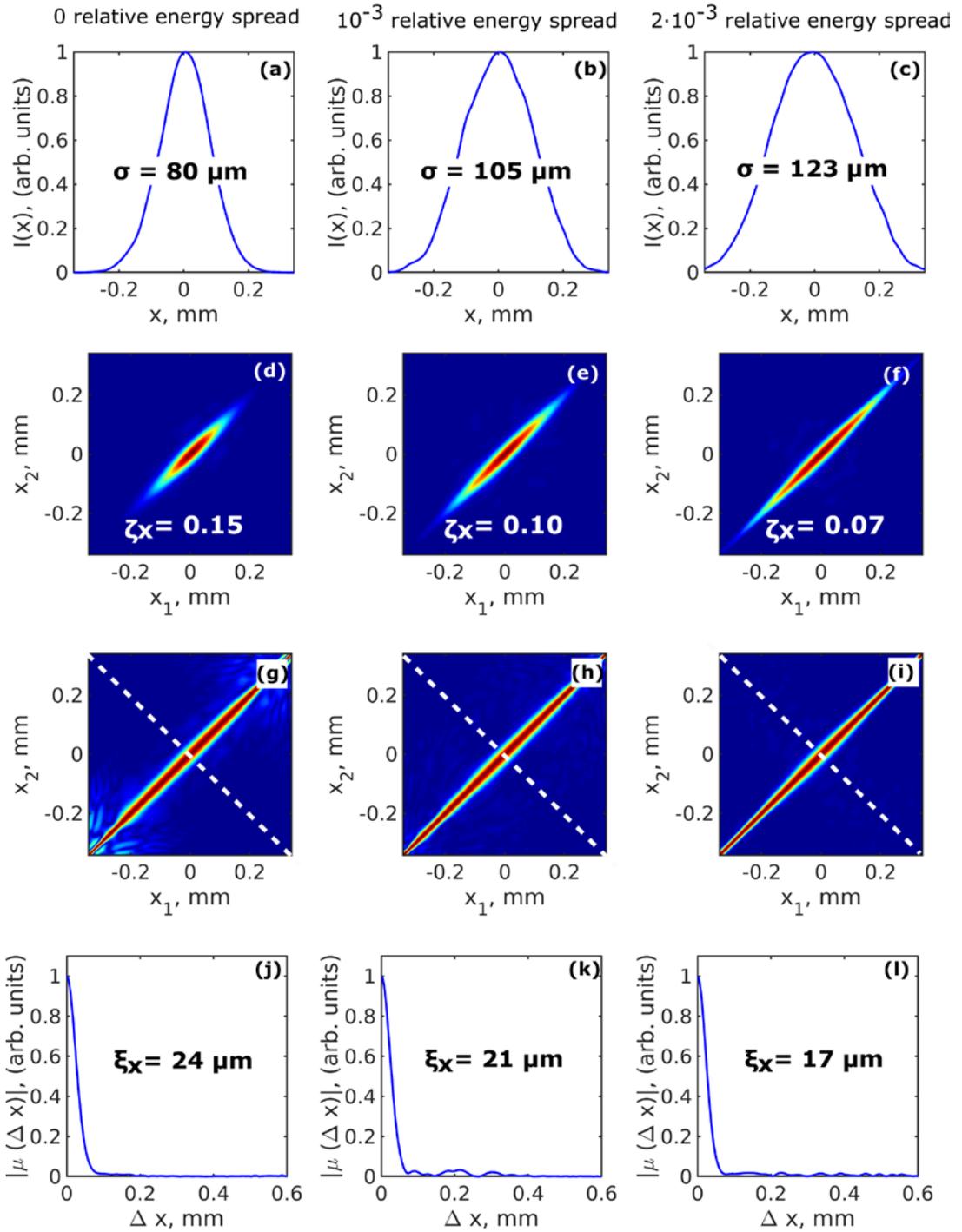

**Figure S4** Simulations of the correlation functions in horizontal direction performed by the XRT software for 50 keV photon energy and different energy spread values. Intensity distribution I($x$) (a-c), absolute value of the CSD in the horizontal direction |W($x_1$,$x_2$)| (d-f), absolute value of the SDC |µ($x_1$,$x_2$)| (g-i), and absolute value of the SDC along the anti-diagonal line (shown in (g-i)) as a function of separation of two points |µ($\Delta x$)| (j-l) simulated in horizontal direction 30 m downstream from the undulator source. In (a-c) $\sigma$ is the rms value of the beam size, in (d-f) $\zeta_x$ is the transverse degree of coherence, in (j-l) $\xi_x$ is the coherence length determined in the horizontal direction.

**500 eV**

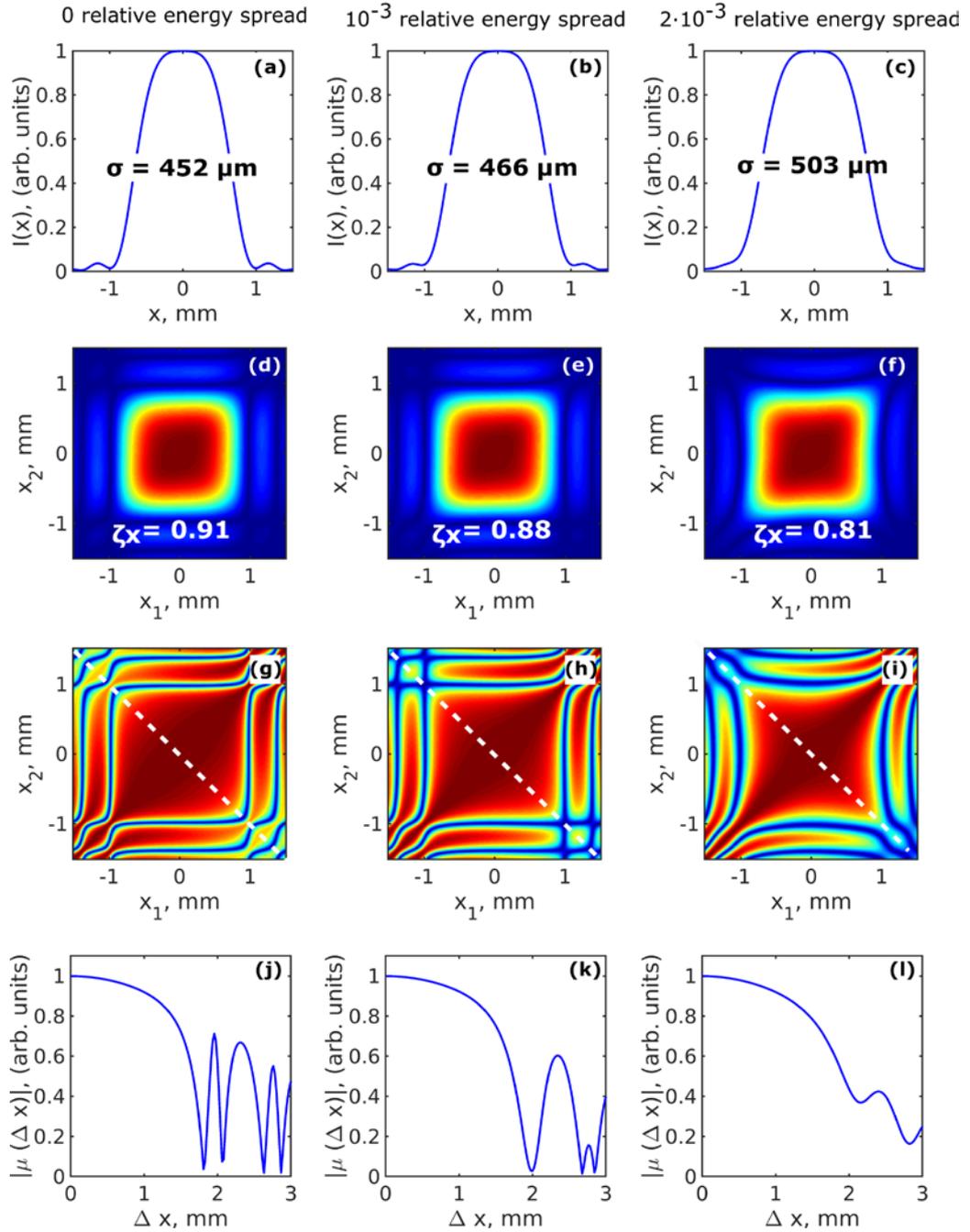

**Figure S5** Simulations of the correlation functions in horizontal direction performed by the analytical approach for 500 eV photon energy and different energy spread values. (a-c) Intensity distribution $I(x)$, (d-f) absolute value of the CSD in the horizontal direction $|W(x_1,x_2)|$, (g-i) absolute value of the SDC $|\mu(x_1,x_2)|$, (j-l) absolute value of the SDC along the anti-diagonal line (shown in (g-i)) as a function of separation of two points $|\mu(\Delta x)|$ simulated in horizontal direction 30 m downstream from the undulator source. In (a-c) $\sigma$ is the rms value of the beam size, in (d-f) $\zeta_x$ is the transverse degree of coherence, in (j-l) $\xi_x$ is the coherence length determined in the horizontal direction.

## 12 keV

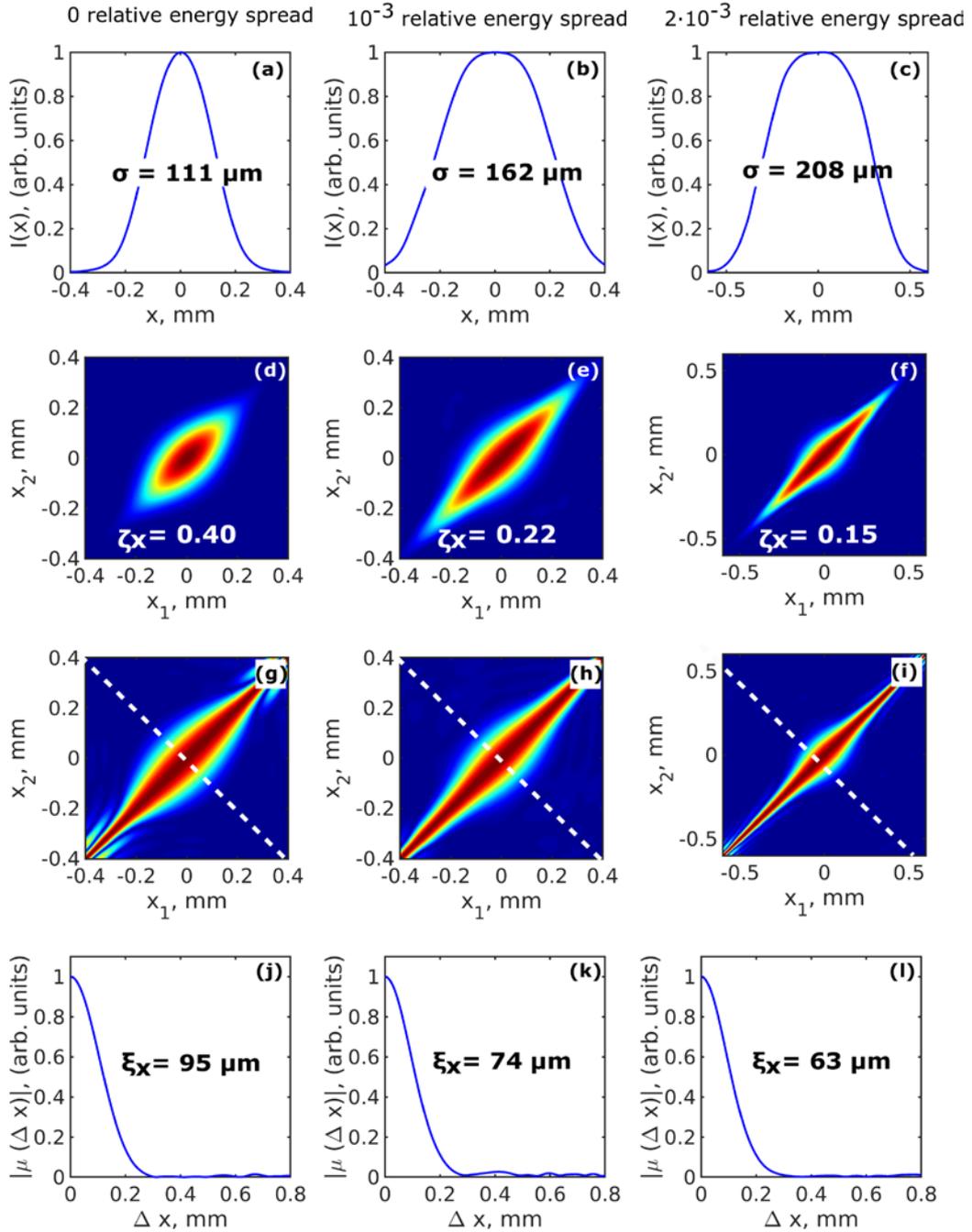

**Figure S6** Simulations of the correlation functions in horizontal direction performed by the analytical approach for 12 keV photon energy and different energy spread values. (a-c) Intensity distribution $I(x)$, (d-f) absolute value of the CSD in the horizontal direction $|W(x_1,x_2)|$, (g-i) absolute value of the SDC $|\mu(x_1,x_2)|$, (j-l) absolute value of the SDC along the anti-diagonal line (shown in (g-i)) as a function of separation of two points $|\mu(\Delta x)|$ simulated in horizontal direction 30 m downstream from the undulator source. In (a-c) $\sigma$ is the rms value of the beam size, in (d-f) $\zeta_x$ is the transverse degree of coherence, in (j-l) $\xi_x$ is the coherence length determined in the horizontal direction.

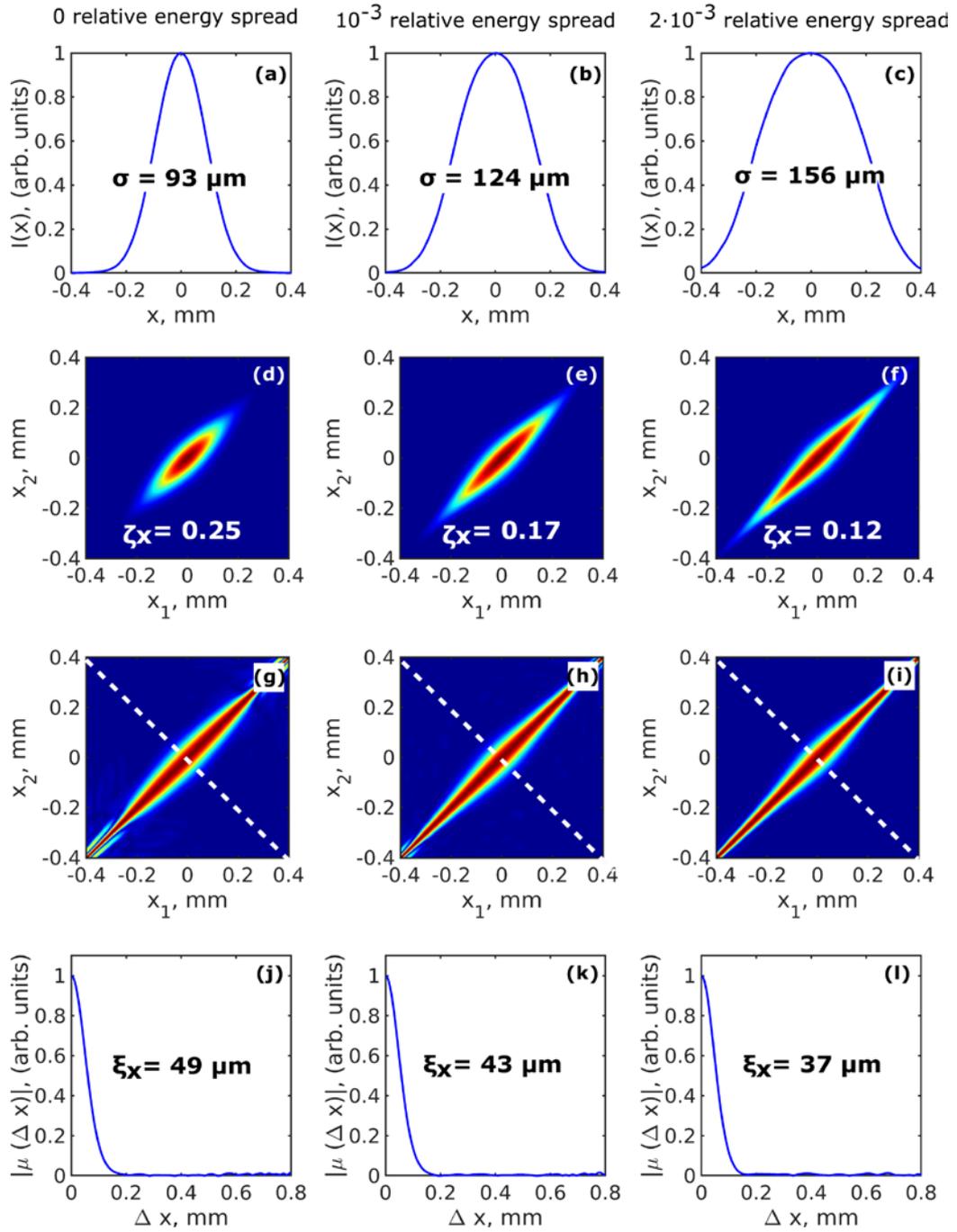

**24 keV**

**Figure S7** Simulations of the correlation functions in horizontal direction performed by the analytical approach for 24 keV photon energy and different energy spread values. (a-c) Intensity distribution $I(x)$, (d-f) absolute value of the CSD in the horizontal direction $|W(x_1,x_2)|$, (g-i) absolute value of the SDC $|\mu(x_1,x_2)|$, (j-l) absolute value of the SDC along the anti-diagonal line (shown in (g-i)) as a function of separation of two points $|\mu(\Delta x)|$ simulated in horizontal direction 30 m downstream from the undulator source. In (a-c) $\sigma$ is the rms value of the beam size, in (d-f) $\zeta_x$ is the transverse degree of coherence, in (j-l) $\xi_x$ is the coherence length determined in the horizontal direction.

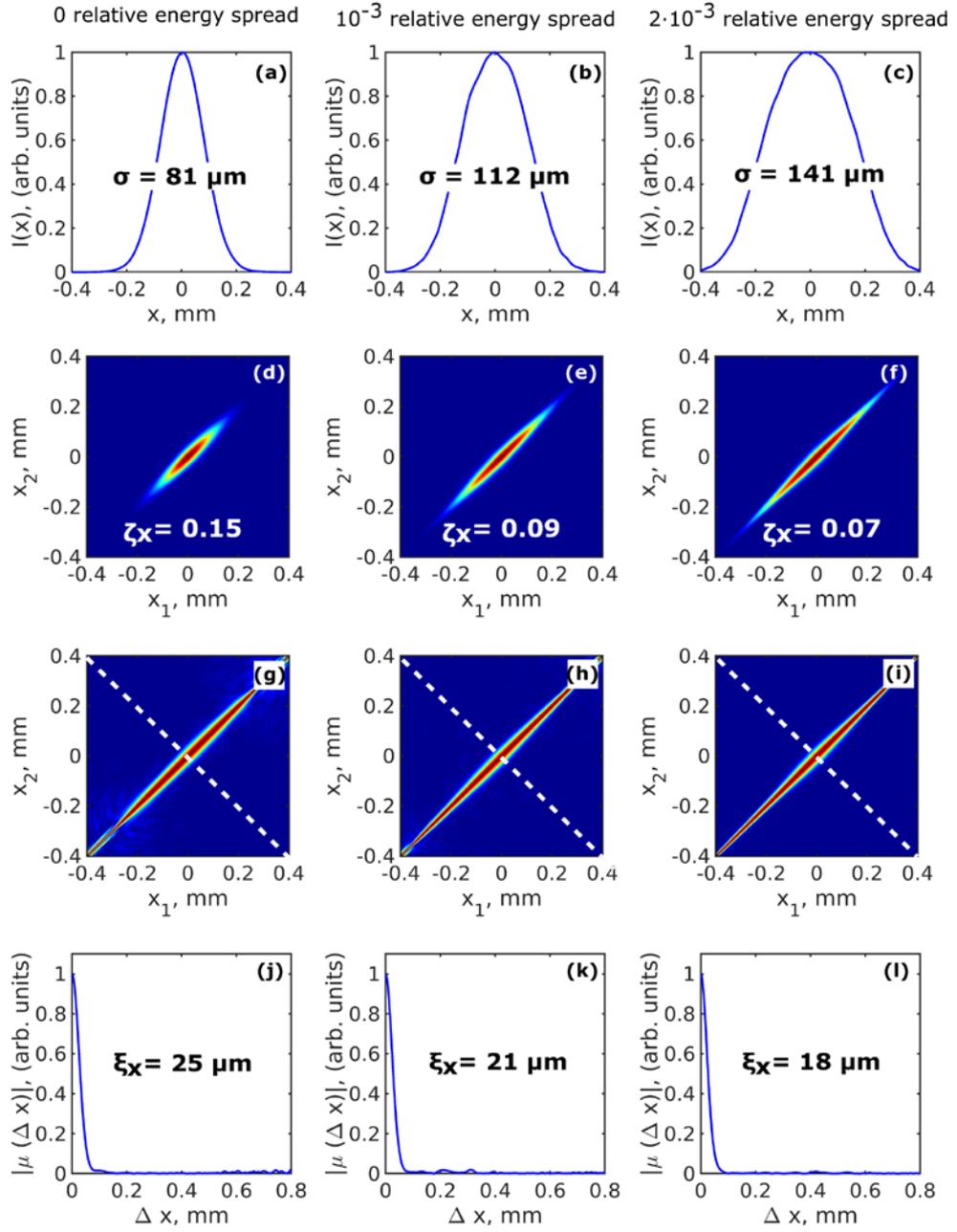

**Figure S8** Simulations of the correlation functions in horizontal direction performed by the analytical approach for 50 keV photon energy and different energy spread values. (a-c) Intensity distribution $I(x)$, (d-f) absolute value of the CSD in the horizontal direction $|W(x_1,x_2)|$, (g-i) absolute value of the SDC $|\mu(x_1,x_2)|$, (j-l) absolute value of the SDC along the anti-diagonal line (shown in (g-i)) as a function of separation of two points $|\mu(\Delta x)|$ simulated in horizontal direction 30 m downstream from the undulator source. In (a-c) $\sigma$ is the rms value of the beam size, in (d-f) $\zeta_x$ is the transverse degree of coherence, in (j-l) $\xi_x$ is the coherence length determined in the horizontal direction.

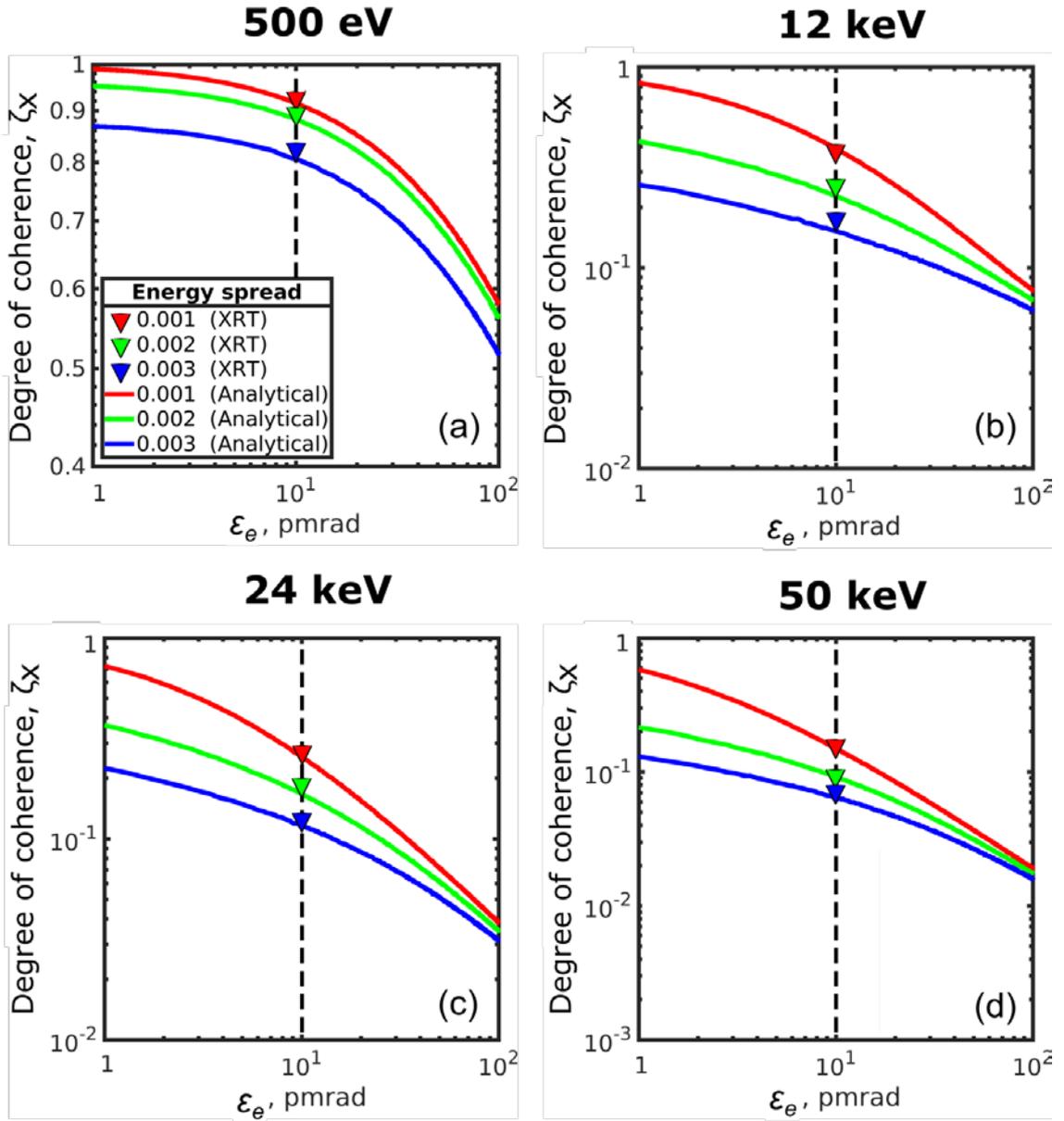

**Figure S9** Simulation of the degree of transverse coherence $\zeta_x$ obtained from XRT and analytical analysis (Eq. (5)) as a function of natural electron emittance for 500 eV, 12 keV, 24keV, and 50 keV.